\documentstyle[11pt,aaspp4]{article}

\newcommand\MSUNYR{\rm M_{\odot}\,yr^{-1}}
\newcommand\MSUN{\rm M_{\odot}}
\newcommand\LSUN{\rm L_{\odot}}
\newcommand\RSUN{\rm R_{\odot}}
\newcommand\Mdot{ \dot{M}}
\newcommand\etal{{\it et al}. }
\newcommand\be {\begin{equation}}
\newcommand\en{\end{equation}}
\newcommand\cm{\rm cm}
\newcommand\din{\rm dyn}

\newcommand\AU{\rm AU}

\newcommand\K{\rm K}

\def\prom#1{\langle #1\rangle}

\begin{document}

\title
{ACCRETION DISKS AROUND YOUNG OBJECTS.  II.
TESTS OF WELL-MIXED MODELS WITH ISM DUST}

\author{Paola D'Alessio \altaffilmark{1}, Nuria Calvet \altaffilmark{2}, Lee Hartmann \altaffilmark{2}}
\author{Susana Lizano \altaffilmark{1}, and Jorge Cant\'o \altaffilmark{1}}
\altaffiltext{1}{Instituto de Astronom\'{\i}a, UNAM, Ap. Postal 70-264, Cd. Universitaria, 04510 M\'exico D.F., M\'exico
Electronic mail: dalessio@astroscu.unam.mx,lizano@astrosmo.unam.mx}

\altaffiltext{2}{Harvard-Smithsonian Center for Astrophysics, 60 Garden St.,
Cambridge, MA 02138, USA;
Electronic mail: ncalvet@cfa.harvard.edu, hartmann@cfa.harvard.edu}

\begin{abstract}
We construct detailed vertical structure models of irradiated accretion disks
around T Tauri stars with interstellar medium dust uniformly mixed
with gas. The dependence of
the structure and emission properties on
mass accretion rate, viscosity parameter, and
disk radius is explored using these models.
The theoretical spectral energy distributions (SEDs)
and images for all inclinations are
compared with observations
of the entire population of Classical T Tauri stars (CTTS) and Class I objects
in Taurus.
In particular, we find that the median near-infrared fluxes can be explained within
the errors with the most recent values for the median accretion rates for CTTS. We
further show that  the majority
of the Class I sources  in Taurus cannot be Class II sources viewed edge-on
because they are too luminous and their colors
would be consistent with disks seen only in a narrow range of inclinations.
Our models appear to be too geometrically thick at large radii,
as suggested by:
(a) larger  far-infrared disk emission than in the typical
SEDs of T Tauri stars;
(b) wider dark dust lanes in the model images than
in the images of HH30 and HK Tau/c; and
(c) larger predicted number of stars extincted by edge-on disks
than consistent with current surveys.
The large thickness of the model is a consequence
of the assumption that dust and gas are well-mixed,
suggesting that some degree of dust settling
may be required to explain the observations.

\end{abstract}


\section{Introduction}
\label{sec_intro}

With the advent of {\it IRAS}, it became clear by the mid-1980s that the infrared
excesses
of many T Tauri stars were probably produced in dusty circumstellar disks rather
than gaseous envelopes (e.g., Rucinski 1985).
However, it was also recognized that
the spectra of the infrared excess of T Tauri stars (Rydgren \& Vrba 1987) was not
of the form
predicted either for a steady accretion disk (Lynden-Bell \& Pringle 1974)
or for a flat disk heated by radiation from the central star (Friedjung 1985),
 with the implication that the outer disk temperatures are
much hotter than predicted by the standard models.
A variety of mechanisms were invoked to explain this result,
but the most robust model has proved to be that of the irradiated ``flared disk''
(Kenyon \& Hartmann 1987, hereafter KH87).  Because the scale height $H$ in
a Keplerian (geometrically thin) disk varies with the sound speed $c_s$ and radius
$R$ as
$H/R \propto c_s R^{1/2}$, any disk temperature distribution which decreases
less rapidly with radius than $R^{-1}$ (expected on quite general grounds)
causes the disk to become proportionately
thicker with increasing $R$.  This ``flaring'' of the disk allows it to absorb
more
radiation from the star, especially at large distances, resulting in spectral
energy distributions (SEDs) that reproduce observations much better
(KH87; Kenyon \& Hartmann 1995, hereafter KH95).

More detailed models of flared disks, which relax the assumption of vertical
isothermality adopted by KH87,
have been constructed by Chiang \& Goldreich (1997, hereafter CG),
who presented a two-isothermal-layer model, to allow for irradiation
heating of the upper layers, and by
D'Alessio et al. (1998, hereafter
DCCL), who 
solve the detailed
vertical hydrostatic disk structure with irradiation and viscous heating
self-consistently.
Although the resulting
model SEDs have been shown to agree with observations of a single T Tauri star
in each of these papers (not the same star in both), they have not been shown to agree with
typical disk emission spectra.  Moreover, there are reasons to suspect that these
models, which assume well-mixed gas and dust throughout the disk,
may not be able to reproduce many results.
In particular, the disk models presented in these
studies are so vertically thick that, if inclinations were randomly distributed,
roughly half of all T Tauri stars should be hidden behind their disks
for typical radii.
Not only would this mean large incompleteness in present surveys of star-forming
regions, it would also suggest
that many of the Class I objects, which were originally thought to be
protostars
with infalling dusty envelopes, might simply be T Tauri stars seen edge-on.
On the other hand, the results of Miyake and Nakagawa (1995) suggested that some T
Tauri
SEDs could be explained with flat disks, possibly implying dust 
settling to the disk midplane. 

In this paper we construct detailed irradiated disk models for T Tauri stars and
compare them
with typical T Tauri SEDs.
This comparison suggests that T Tauri disks are thinner than
predicted by the
models with complete mixing of ISM dust and gas.  Survey incompleteness
due to stars hidden behind their edge-on disks does not seem to be very large.
We also argue that most Class I objects in Taurus cannot simply be edge-on disks,
but are probably
protostars surrounded by their infalling envelopes, as previously proposed (Adams,
Lada, \& Shu 1987; Kenyon, Calvet \& Hartmann 1993).
The models suggest that the assumptions of complete mixing of dust and gas
and standard ISM dust may need to be relaxed, and we explore such new models
in a subsequent paper.

\section{Disk Models}
\label{sec_model}

\subsection{Assumptions and Methods}
\label{sec_method}

We solve self-consistently the complete set of vertical structure
equations, including irradiation and viscous heating, resulting
in detailed profiles of temperature and density with
vertical height.
The model of the disk vertical structure  is calculated with the equations and
method described by
DCCL, as summarized briefly here.
The disk is assumed to be geometrically thin and
in steady state with a constant mass accretion rate $\Mdot$, and in
vertical hydrostatic equilibrium. The
viscosity is described using the $\alpha$ prescription (Shakura \& Sunyaev 1972),
with a
viscosity coefficient written as
$\nu_t = \alpha c_s H$, where $c_s$ is the local sound speed, $H$ is the local
scale
height of the gas and $\alpha$ is the viscosity parameter, assumed
constant through the disk. The scale height is given by $H=c_s(T)/\Omega$,
where $\Omega$ is the angular Keplerian velocity and $T$ is the local temperature.
A characteristic scale height of the
disk is $H_c$, calculated with the sound speed evaluated at the midplane
temperature $T_c$.
The disk has azimuthal symmetry, and its equations are written in cylindrical
coordinates $(R,z)$, where $R$ is the radial distance from the star, in a
direction parallel to the disk
midplane, and $z$ is the vertical distance from the midplane, in a direction
parallel to the
rotation axis.

The radiation field is separated into two frequency
ranges as proposed by Calvet et al. (1991, 1992),
one characteristic of the stellar effective temperature and
the other corresponding to the local disk temperature.
We use mean opacities to quantify the disk interaction
with the two  different
radiation fields.
The Planck mean absorption coefficient for a stellar effective temperature
$T_*=4000 \ \K$
is similar to the monochromatic absorption coefficient at $\lambda = 1 \ \mu
m$ (e.g. KH87).
The stellar radiation field is described as a beam of parallel rays
impinging on the disk surface at an angle $\cos^{-1} \ \mu_0$ relative
to the disk normal.

The differential equations for the disk vertical structure are
the same as those in DCCL. We write them here in a slightly different and clearer
way.
First, we consider the equations for radiative transfer in the ``disk-frequency
range''
(Mihalas 1978),
\be
{dF_{d} \over dz}=4 \pi \kappa_P \rho \biggl ( {\sigma_R T^4 \over \pi} -J_d
\biggr ) \ \ , \label{eq:fd}
\en
\be
{dJ_d \over dz} = - {3 \over 4 \pi} \ \chi_R \rho F_{d} \ \ , \label{eq:jd}
\en
where $F_d$ and $J_d$ are the  flux and the mean intensity, respectively,
at the disk frequency range, $\rho$ is the mass density,  $\kappa_P$ and $\chi_R$
are the Planck
mean and the Rosseland mean opacities, respectively, and $\sigma_R$ is the
Stefan-Boltzmann constant.

The disk energy balance equation is
\be
{dF \over dz} = \Gamma_{vis} + \Gamma_{ion} +\Gamma_{irr} \ \ , \label{eq:dfdz}
\en
where $F$ is the total flux given by the sum
of the radiative, convective, and turbulent energy fluxes (DCCL).
Here, $\Gamma_{vis}$ is the heating due to viscous dissipation (e.g., Frank et al.
1992),
$\Gamma_{ion}$ is the heating due to ionization by cosmic rays and radioactive
decay
(Nakano \& Umebayashi 1986; Stepinski 1992) and $\Gamma_{irr}$ is the heating by
irradiation.
This last heating function can be written as (see DCCL):
\be
\Gamma_{irr} = 4 \pi \kappa_P^* \rho [ J_s + J_i] \ \ ,
\label{eq_gammairr}
\en
where $\kappa_P^*$ is a mean opacity calculated using the
Planck function at the stellar effective temperature $T_*$ as a weighting
function,
and $J_i$ and $J_s$ are the mean intensities of the direct and scattered stellar
radiation field.
To calculate the transfer of the
stellar radiation through the disk, we use the formulation of
Calvet et al. (1991, 1992), with the assumption of constant
albedo in the region of the atmosphere where most of the stellar radiation is
scattered.
The mean intensity of the direct stellar radiation field  is given by
\be
J_i = J_{irr}  e^{-\tau_{s,z}/\mu_0} \ \ ,
\label{eq_jotai}
\en
where $\tau_{s,z}$ is the vertical optical depth to the stellar radiation
and $J_{irr}$ is the mean intensity of the stellar radiation at
the top of the disk.  The scattered mean intensity is
\be
J_s = {s \ F_{irr} (2 +3 \mu_0) \over 4 \pi [ 1 + (2 g/3)](1-g^2 \mu_0^2)} \ e^{-g
\tau_{s,z}}
- {3 \mu_0 s F_{irr} \over 4 \pi (1-g^2 \mu_0^2)} e^{-\tau_{s,z}/\mu_0} \ \ ,
\label{eq_jotas}
\en
\noindent
where $s$ is the mean albedo, $g={[3(1-s)]^{1/2}}$, $F_{irr}$
is  the  stellar flux intercepted by the ``irradiation surface'' of the disk.
We take $\mu_0$ as the cosine of the angle between the incident direction and
the normal to this surface.
The irradiation surface $z_s(R)$ is defined
by requiring that the optical depth in the direction to the star is $\tau_s=2/3$,
with $\tau_s$ calculated integrating the opacity in the radial direction, i.e.,
\be
\tau_s = \int_{r_{min}}^r \ \ \chi_s(R,z) \ \rho(R,z) \ dr \ \ ,
\en
where $r = {(z^2 + R^2)^{1/2}}$, and $\chi_s$ is a mean opacity calculated
including true absorption
and scattering, using the Planck function evaluated at $T_*$ as the weighting
function.
Since this optical depth depends on the disk
structure, which in turn depends on the irradiation flux,
we iterate to find the self-consistent disk structure.
Notice that in equations (\ref{eq_jotai}) and (\ref{eq_jotas}) we have used the plane parallel
approximation, so $\tau_s \approx \tau_{s,z}/\mu_0$, but this approximation is not used to find
the height of the irradiation surface.

The turbulent and/or convective energy transport are included as
\be
{dT \over dz} = - \nabla(F_{rad},F,T,P) \ {T \over P} \ g_z \ \rho \ \ ,
\label{eq4}
\en
where $\nabla$ is calculated taking into account that the disk may have a finite
optical depth
 (i.e.,
without using the diffusion approximation) and that the convective elements lose
energy by
radiation and by the turbulent flux (see DCCL and D'Alessio (1996) for details).

Finally, we have the vertical hydrostatic equilibrium equation
\be {dP \over dz} \
 = - \rho g_z  \ \ ,
\label{eq5}
\en
where $P$ is the gas pressure, $g_z$ is the $z$ component of the stellar gravity,
i.e.,
$g_z= G M_* z/(R^2+z^2)^{3/2}$,  where $M_*$ is the stellar
mass, and we have
neglected radiation pressure and the disk self gravity.

The boundary conditions for these equations are basically the same as in
DCCL. At the disk midplane the fluxes are zero because of reflection symmetry.
The height of the disk $z_\infty$ is an unknown boundary
where we specify the ambient gas pressure $P_\infty$, the emergent
fluxes produced by viscous dissipation $F_{vis}$, ionization heating $F_{ion}$ (by
cosmic rays and radioactive
decay) and irradiation $F_{irr}$, and the mean radiative intensity
at the disk frequency
$J_d = h F_{irr}$, where $h$ is a function of $s$ and $\mu_0$ (cf. DCCL).
The turbulent and convective energy fluxes equal zero at
$z_\infty$ by definition.
We assume $P_\infty=10^{-10} \ \din  \ cm^{-2}$, which is an arbitrary value,
but small enough that its precise value does not
significantly affect the resulting disk structure.

The intercepted irradiation flux $F_{irr}$ is calculated as the component of the
incoming stellar
radiant energy normal to the irradiation surface $z_s$ (instead of
the surface where the pressure is the ambient pressure $z_\infty$, as in DCCL).
Note that the normal to the disk surface is not precisely in the z direction.
Since the slope of this irradiation surface is used to
evaluate $F_{irr}$ and $\mu_0$ (see KH87) we
estimate this slope using the following procedure. We
fit  a power law to the height of the irradiation surface,
$z_s(R) \sim R^b$, in an interval $\Delta log R \sim 0.55$
centered at each radius and the slope is estimated as $d z_s/dR \approx b z_s/R$,
using the actual value of $z_s(R)$. This smoothing procedure accelerates the
convergence and
avoids numerical instabilities, and is justified
since irregularities in the shape of the disk surface are damped in times
shorter than the viscous timescale (Cunningham 1976; D'Alessio et al. 1999).
The irradiation flux is the fraction of the intercepted stellar flux
absorbed or scattered by the disk. We estimate this fraction as
$f=(1-e^{-\tau_{s,total}})$,
where $\tau_{s,total}$ is the total optical depth of the disk
to the stellar radiation.
We calculate the flux as described by
KH87, but multiply by $f$ to take into
account that there can be optically thin regions, which cannot absorb all the
intercepted stellar radiation. For typical
parameters of T Tauri disks, $f \sim 1$.

The disk dust opacity is calculated using optical constants from Draine \& Lee
(1984) and
the MRN size distribution  (Mathis, Rumpl \& Nordsieck 1977). For the gaseous
component, we
use the opacities described by Calvet \etal (1991). In this paper, we
assume that gas and dust are well mixed and thermally coupled.

\subsection{Fiducial Model}
\label{sec_estruc}

To illustrate the basic results of the detailed disk calculations,
we constructed a fiducial disk model using typical parameters of CTTS.
The central star is assumed to have a radius $R_*=2 \ \RSUN$,
a mass $M_*=0.5 \ \MSUN$, and an effective temperature $T_*=4000$ K
typical of K7-M0 T Tauri stars in Taurus (KH95).
We adopt a mass accretion rate $\Mdot=10^{-8} \ \MSUNYR$ and a viscosity
parameter $\alpha=0.01$, as suggested by recent studies
(Gullbring et al. 1998; Hartmann et al. 1998).
Variations of the outer disk radius are easily incorporated in our
formalism; the fiducial model has $R_d=100 \ \AU$.
We also include
an inner disk hole, consistent with a magnetospheric
radius of $R_{hole} = 3 \ R_*$ (Kenyon et al. 1994a;
Kenyon, Yi, \& Hartmann 1996; Meyer, Calvet, \& Hillenbrand
1997).

\begin{figure}
\plotone{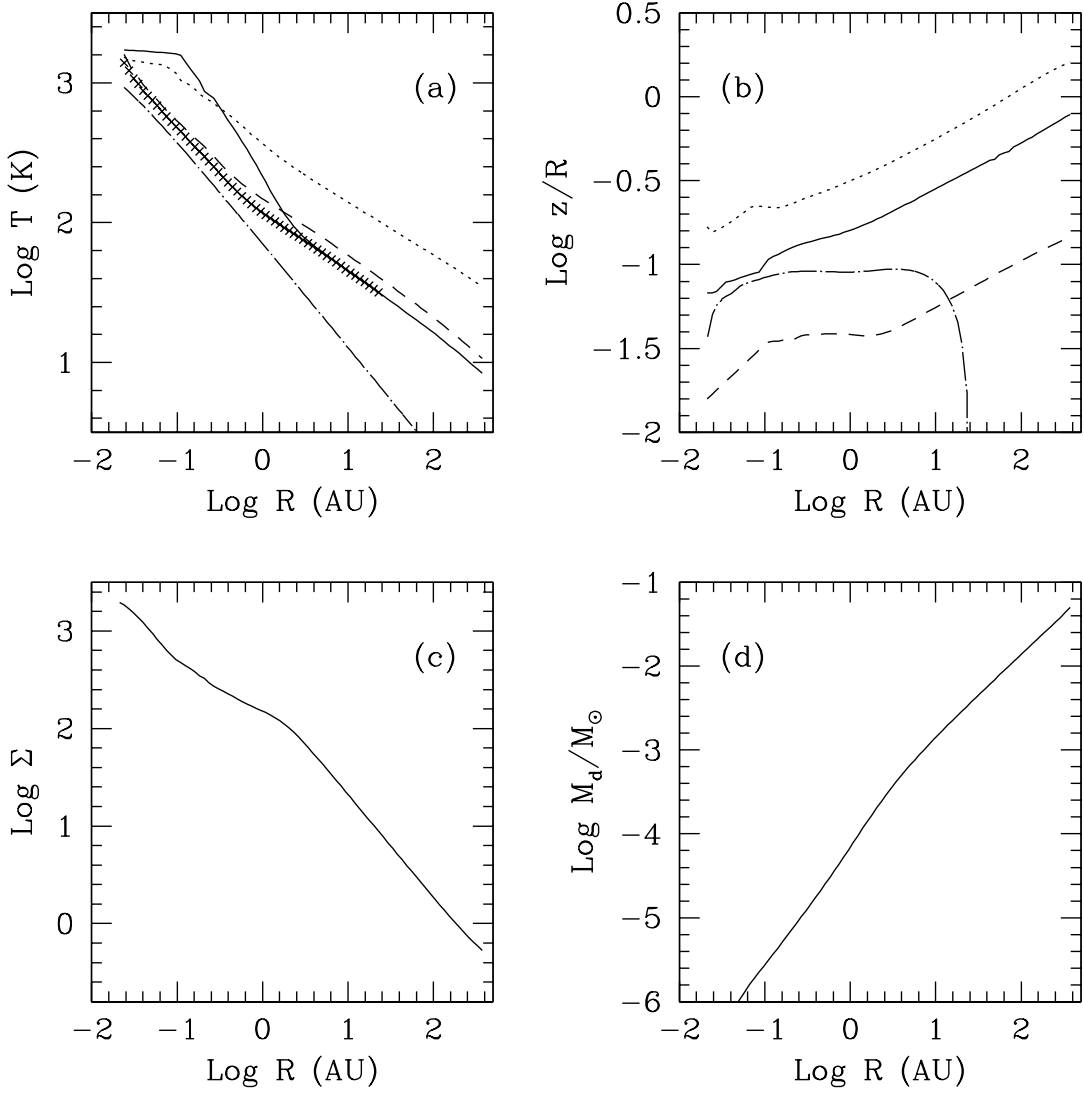}
\caption{Structure of the fiducial model. Panel (a)-
characteristic temperatures: $T_c$ (solid), $T_0 = T(z=z_s)$ (dotted),
 $T_{irr}$ (short-dashes), $T_{vis}$ (dots and long dashes)
and $T_{phot}$ (crosses). Panel (b)-  Characteristic heights: $z_\infty/R$
(dotted),
$z_s/R$ (solid), $H_c/R$ (dashed) and $z_{phot}/R$ (dot-dashed). Panel (c) - Mass
surface
density. Panel (d) - Cumulative mass of the disk as a function of distance to the
central star. }
\label{fig_estruc}
\end{figure}

Figure \ref{fig_estruc} shows characteristic temperatures, heights,
surface density, and
mass of the fiducial disk model as functions of the distance to the central
star.
The plotted temperatures are the temperature at the midplane $T_c$, the
irradiation temperature $T_{irr}$, which is the effective temperature
corresponding to the irradiation flux $F_{irr}$,  the surface temperature $T_0$,
given by the temperature at $z=z_s$,
the viscous temperature $T_{vis}$, defined as ${\sigma}_R T_{vis}^4 = F_{vis}$, and
the photospheric temperature $T_{phot}$, defined as the temperature at the height
where the Rosseland mean optical depth is $\tau_R (z)=2/3$, so it
is only plotted for those regions where the disk total vertical Rosseland optical
depth is larger than this value.

A comparison of $T_{phot}$ and $T_{vis}$ indicates that for $R \gtrsim 1 \ \AU$
irradiation is the main heating agent of the disk upper layers,
while in the inner disk the viscous flux is significant in comparison with the
irradiation flux.
In this region, $T_c$ is larger
than $T_{phot}$ because the disk is optically-thick to its own radiation, resulting
in effective trapping of the viscously-generated flux.
The plateau in $T_c(R) \sim 1600 \ \K$
is a result
of dust sublimation, which acts as a kind of thermostat
because the gas opacity for $T \lesssim 2000$~K is very low.
As the central temperature rises beyond $1600$~K, the dust opacity which produces
the high central temperature by trapping of viscous energy disappears.
This results in sublimation of dust near the midplane, while the layers
near the photosphere
retain enough dust opacity to produce a nearly fixed central temperature.

The total optical depth of the fiducial disk model
at its own radiation $\tau_R$ becomes $\lesssim 10$
for $R \gtrsim 5 \ \AU$. Beyond $R \sim 30 \ \AU$, it is smaller than 2/3
and $T_{phot}$ is no longer defined;
however, the central temperature
dependence on radius is similar to that of the regions
where the disk is optically thick to its own radiation,
as long as the optical depth to the stellar radiation
is very large.  To see the reason for
this, consider the limit where the cosmic ray and viscous heating can
be neglected. In this limit, the energy balance equation 
(eq. \ref{eq:fd}) can be written as
\be
4 \pi \kappa_P \rho \biggl ( {\sigma_R T^4 \over \pi} -J_d
\biggr ) \ \ ~=~ \Gamma_{irr}\,. \label{eq:first}
\en
using $F \sim F_d$ and equation (\ref{eq:dfdz}).
When the disk is very optically thick to the stellar radiation,
the high optical depth prevents
the stellar radiation from reaching directly the disk midplane
and
the irradiation heating $\Gamma_{irr} \rightarrow 0$,
(eq. \ref{eq_gammairr}), so the
temperature then depends only upon $J_d$. Similarly,
near the midplane, $F_d \sim 0$ (by reflection symmetry)
and equation (\ref{eq:jd}) yields $J_d \sim J_d(z_{\infty})$.
So, $J_d$ is approximately constant in height at given radius; using
the boundary condition, it is given by
\be
J_d \sim h F_{irr}\,,
\en
where $h$ is the relevant Eddington factor (cf. \S \ref{sec_method}).
Thus energy balance yields
a central temperature
\begin{equation}
\sigma T^4 \sim \pi h F_{irr}
\end{equation}
and therefore the temperature has the radial dependence of the fourth root of the
irradiation flux, in this case $T_c \approx R^{-3/7}$.

The terms on the left-hand side of the energy balance
equation (\ref{eq:first}) represent integrals over frequency,
and in a more complete treatment
can be written as $4 \pi \rho (\kappa_P B - \kappa_J J_d)$,
where $\kappa_J$ is the mean opacity weighted by the disk mean intensity.
We are assuming that $\kappa_P = \kappa_J$, i.e. that
the disk radiation field has a spectrum comparable to that of the
local Planck function $B$.  
The characteristic temperature of the hotter upper regions where most of the
stellar radiation is absorbed, and therefore where most of $J_d$ is produced
is roughly $T_{irr}$.
Figure 1a demonstrates that $T_{irr}$ is only slightly higher than $T_c$ from the
detailed numerical calculations, so it appears to be a reasonable approximation to set
$\kappa_P = \kappa_J$ in our models.

Other treatments attempt to include the difference in effective wavelength
of the radiation emitted in different parts of the disk in an approximate way,
but these approaches have other difficulties.
For instance, CG assumed that the hot upper layer which absorbs the central
star's light radiates into the disk at the maximum temperature, 
so that $J_d$ can be characterized by a temperature $\sim T(z_\infty)$, and this 
is much hotter than the central temperature.  However, this approximation 
overestimates the characteristic temperature of $J_d$ because most of the 
radiation of the upper layers is produced at lower temperatures than $T(z_\infty)$,
as discussed above.  
Moreover, the CG treatment differs most from ours in the case where the
disk is assumed to be optically thick to the hot layer radiation but optically thin to its 
own radiation; but CG ignore the vertical stratification 
of the radiation field, and for optically-thick transfer, the hot layer radiation
will be degraded into lower-temperature emission as the energy
moves toward the disk midplane.

In the case where the central layers of the disk are
optically thick to the radiation impinging upon them
but optically thin to their own radiation, a typical assumption
is to write the energy balance equation as
\be 
F_{irr} = 4 \tau_p \sigma_R {T}^4\,
\label{eq_tconst}
\en
namely, optically thick
heating is balanced by optically thin cooling. 
The temperature of the assumed isothermal layer
obtained from this equation has a much
flatter radial dependence than $R^{-3/7}$ (Cant\'o et al. 
1995; Aikawa et al. 1999).
We can obtain this equation by taking
the integral
over the  vertical optical depth of the  energy
transport equation (\ref{eq:fd}), from which we can write
\be
F_d \sim 4 \tau_P \sigma_R \prom{T}^4   - 4 \pi \tau_P J_d
\en
where
\be
\prom{T}^4 = {1 \over \tau_P} \int_0^{\tau_P} {T(\tau_P^\prime)}^4 d \tau_P^\prime \ \ .
\label{eq_tmean}
\en
\noindent
With $F_d \sim F_{irr}$, this equation becomes
\be
F_{irr} (1 + 4 \pi  h \tau_P) \approx F_{irr} \approx 
{4 \tau_P \sigma_R \prom{T}^4 }  \ \ 
\label{eq_firrmean}
\en
for $\tau_P << 1$,
which is similar to equation (\ref{eq_tconst}). 
This shows that the temperature that is obtained by using
equation (\ref{eq_tconst}) is a mean temperature over the vertical 
structure and not the actual
physical depth-dependent temperature of the interior.
This distinction may be important when considering disk
properties which are very very sensitive
to local conditions  as for instance,
the molecular processes which determine
the chemical composition of the disk (Aikawa et al. 1997, 1999).

Irradiation, enhanced by the disk curvature, also dominates the thermal structure
of the disk atmosphere. The surface temperature is higher than the
photospheric temperature, since the stellar incident energy is mostly deposited in the upper
layers of the disk; this, in turn, is due to the large dust opacity at wavelengths characteristic
of the stellar radiation and to the oblique entry of this radiation into
the disk (Calvet \etal 1991, 1992; Malbet \& Bertout 1992; CG).
This is the temperature inversion
found by Calvet et al. (1991), which can produce molecular and
silicate bands in emission even when the disk is optically thick.
The inversion is present even in the
regions where the disk is optically thin to its own radiation, so $T_c < T_0$
(Figure \ref{fig_estruc}a).

Figure \ref{fig_estruc}b shows some characteristic heights of
the disk as a function of radius. We have plotted  the irradiation
surface height $z_s$, the photospheric height $z_{phot}$ (where
$\tau_R=2/3$), the maximum height of the disk $z_\infty$ (where $P=P_\infty$, see
\S \ref{sec_method}) and
the gas scale height $H_c$, calculated using the sound speed at the disk midplane.
In particular, $z_s$ can be taken as proportional to $H_c$ for $R \gtrsim  4 \
\AU$,  as in the
approximation usually found in the literature, with a proportionality constant
$\sim 5$.
The height $z_s$ corresponds to a flared surface,
 with $d z_s/dR > z_s/R$,  at least for $R \lesssim 340 \ \AU$ which
is the maximum disk radius we have considered.
On the other hand, the photospheric height has a maximum beyond which it
decreases as the disk becomes optically thin to its own radiation.  However,
since the disk is still optically thick to the
stellar radiation in the radial direction,  $z_s(R)$ is flared
in regions where $z_{phot}$ bends down. The behavior of $z_{phot}$ and
the bump shown by $H_c$ led Bell \etal (1997) to conclude that the outer disk
was in the
shadow of the inner disk, ignoring the role
of the material  at the upper disk atmosphere, which is able to
absorb and scatter stellar radiation effectively.
Our results contradict their claim, since a more realistic treatment, allowing
for the very different wavelengths at which light is absorbed and emitted by the disk,
results in no shadowed regions.

The surface density of mass $\Sigma$ is the integral along the vertical direction of the
volumetric mass density $\rho(R,z)$.
In steady disk models using the $\alpha$ prescription,
$\Sigma \propto \Mdot/\nu_t \propto \Mdot \Omega / \alpha T$
(e.g., Frank et al. 1992).
With $T \sim T_c \propto R^{-1/2}$ in the outer disk, $\Sigma \propto R^{-1}$,
as shown in Figure \ref{fig_estruc}c.
The surface density flattens towards smaller radii, where
the midplane temperature increases in the optically thick annuli of the disk.
The mass of the disk for the assumed viscosity parameter and mass accretion rate
is
$M_d \sim 0.016 (R_d/{\rm 100 AU})\ \MSUN$
for a disk radius in the range $1 \ \lesssim R_d \lesssim 300 \ {\rm AU}$
(Figure \ref{fig_estruc}d), which
is similar to
the typical masses estimated from mm observations
of CTTS (Beckwith et al. 1990; Osterloh \& Beckwith 1995).

\subsection{Consistency of the plane-parallel approximation}
\label{sec_consistency}

The incident stellar flux is calculated assuming there
is a well-defined surface $z_s$ into which the radiation
enters.  Given the irradiation flux, we solve the transfer of stellar
radiation, assuming a plane parallel geometry.
This may not be a good assumption for the outer regions
of the disk, where the height of the irradiation surface becomes of the order
of the radial distance.
To test this approximation, we have used the two-dimensional disk
structure to calculate {\it a posteriori} the mean intensity of
the direct stellar radiation at each depth $J_i(R,z)$
attenuated by the radial optical depth,
\be
J_i = {\sigma_R T_*^4 \over \pi} \  {1 \over 4 \pi} \int_{\Omega_*} \exp(-\tau_s)
d \Omega \ \ ,
\en
where the integration is performed over the solid angle subtended by the star, as
seen
from each point of the disk. The stellar disk is divided into bands parallel to
the disk midplane
and symmetrically
distributed with respect to the line between the center of the star and a given point
of the disk.
Ignoring the stellar diffuse field,
the local heating by stellar radiation can be evaluated as
\be
\Gamma_{irr,i}(R,z) = 4 \pi \kappa_P^* \ \rho J_i \ \ ,
\en
where the subindex $i$ refers to the direct incident stellar radiation.
Thus, the flux corresponding to the irradiation input of energy,
which has to emerge at the top of the disk is

\be
F_i(z_\infty) = \int_0^{z_\infty} \ \Gamma_{irr,i}(R,z) dz \ \ .
\label{eq_firr2d}
\en

\begin{figure}
\plotone{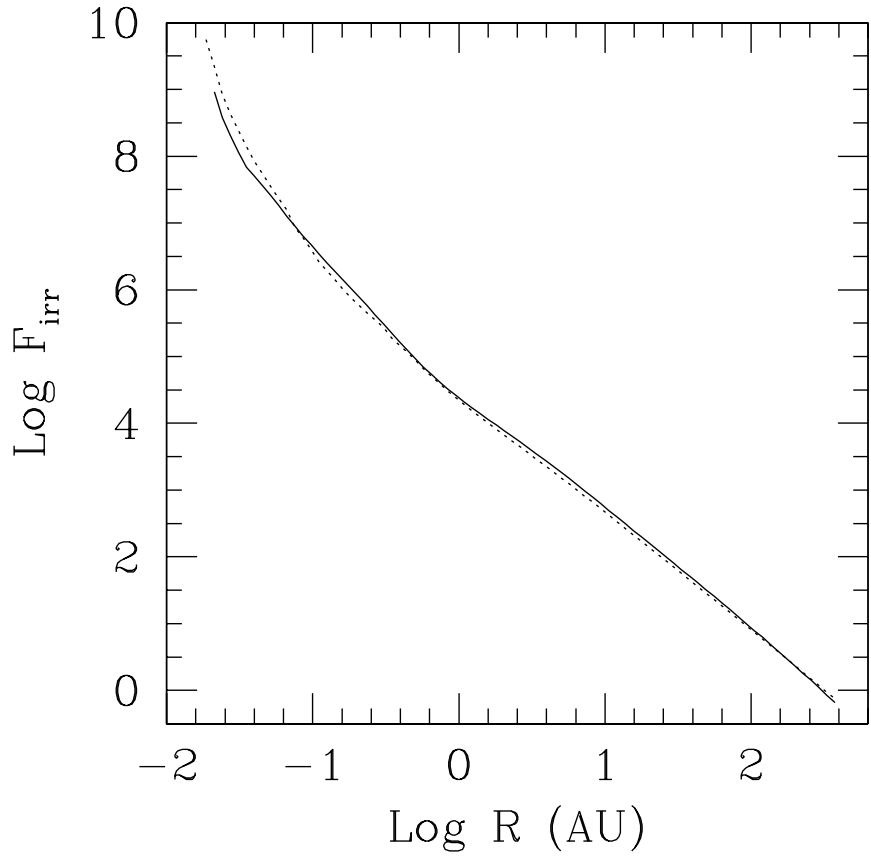}
\caption{Comparison between the irradiation flux from the
plane-parallel calculation (solid line) and $F_{i}(z_\infty)$ given by a ray by
ray integration of the
transfer equation of the stellar radiation (dotted line).}
\label{fig_firr}
\end{figure}

Figure \ref{fig_firr} shows the flux $F_i(z_\infty)$
and $F_{irr}$ given by the stellar flux intercepted by the irradiation surface,
absorbed
by the disk,  calculated
using the plane-parallel approximation and the iterative method described in \S
\ref{sec_method}.
To make a consistent comparison, the flux $F_i(z_\infty)$ is smoothed fitting a
local power law in intervals of $\Delta log R = 0.55$  centered at each radius,
which is the same kind of smoothing procedure we have done in the iterative plane
parallel
calculation.  Figure \ref{fig_firr} shows that both fluxes are very similar.
Since one of them results  from a ray by ray integration of the
transfer equation of the stellar radiation, and the other
was obtained using the plane-parallel approximation, we conclude that the
latter is a good approximation to describe the effect of the stellar irradiation
on the disk.

The plane-parallel approximation breaks down for the transfer of radiation in the
``disk-frequency'' range in the outer disk, where the relevant optical depths can
become
small.  We estimate that this does not greatly affect our calculated temperature
distributions,
which are effectively determined by the balance between the irradiation flux heating,
which
is relatively well known as discussed above, and optically-thin
cooling, which
is insensitive to geometry.  A detailed test of this approximation requires
two-dimensional
radiative transfer and is beyond the scope of this paper.

\subsection{The $\Mdot-\alpha$ Parameter Space}
\label{sec_magrid}

\begin{figure}
\plotone{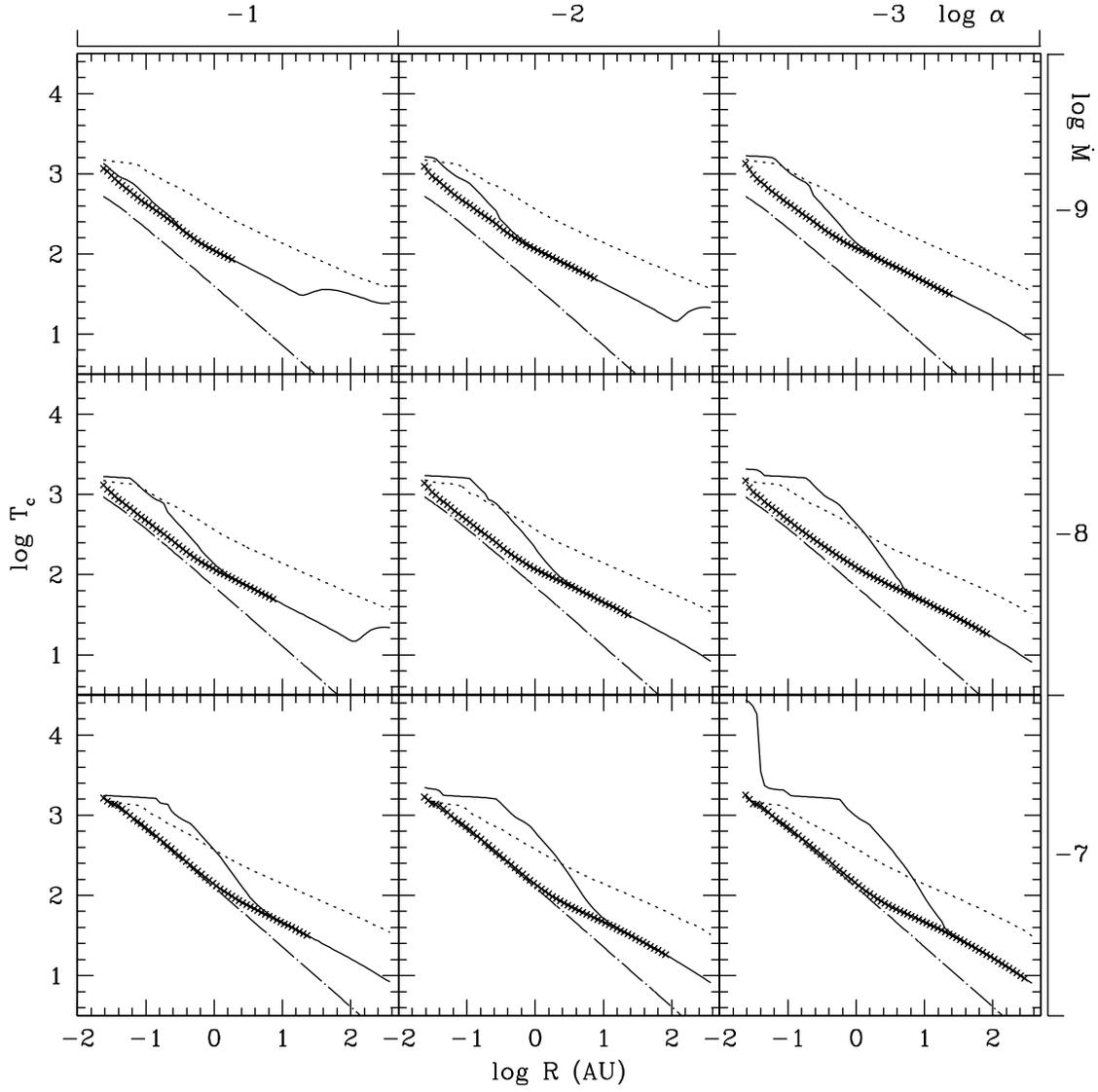}
\caption{Temperatures of disk models with $\log (\Mdot/ \MSUNYR)=-9, \ -8, \ -7$,
$\log \alpha=-3, \ -2, \ -1$, around a central star with $M_*=0.5 \ \MSUN$, $R_*=2
\ \RSUN$ and
$T_*=4000 \ \K$. The plotted temperatures are: $T_c$ (solid line), $T_0$
(dotted line),
$T_{phot}$ (crosses) and $T_{vis}$ (dot-dashed line).
}
\label{fig_tempgrid}
\end{figure}

\begin{figure}
\plotone{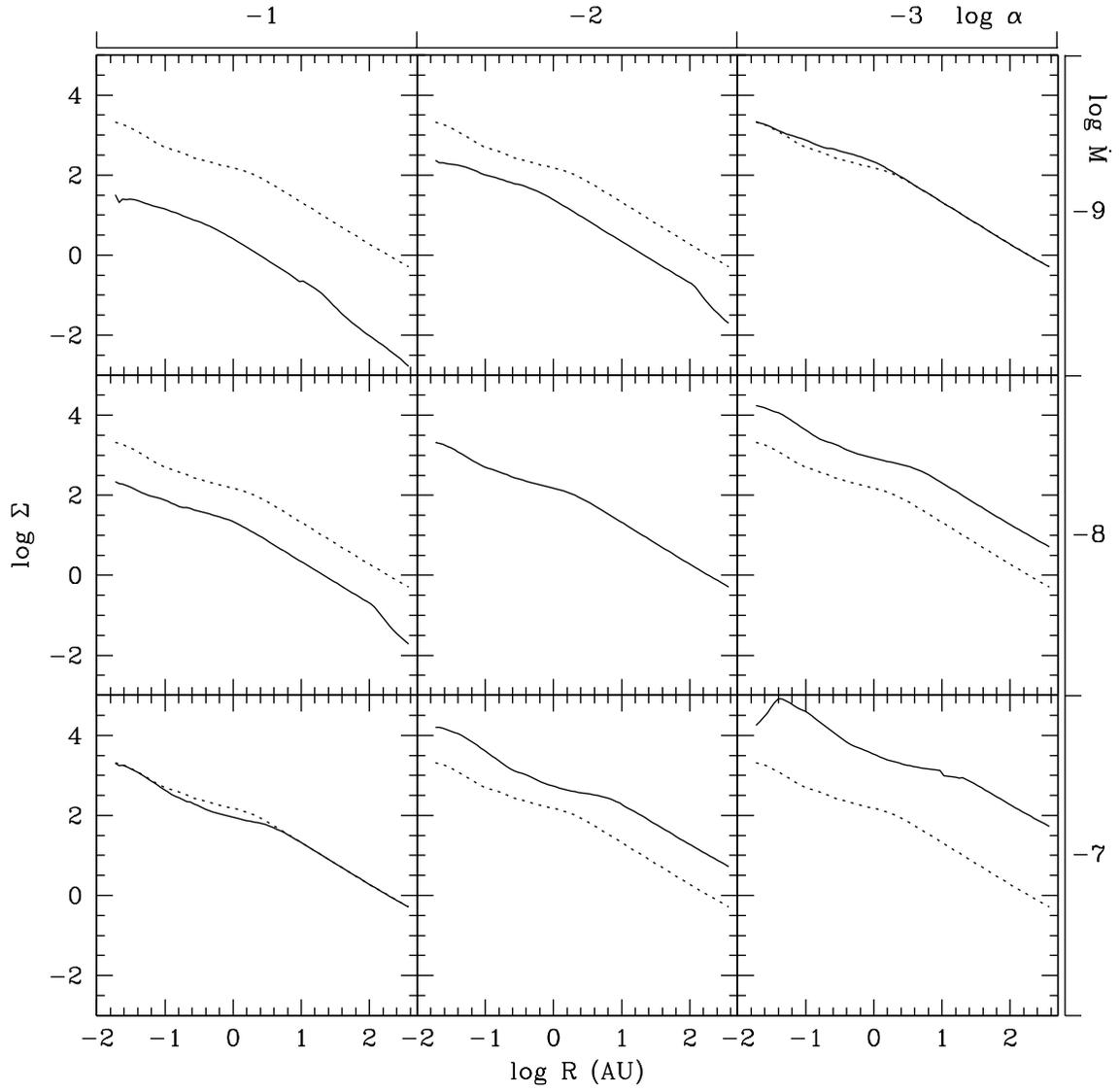}
\caption{Mass surface density  of the disk models plotted in Figure
\ref{fig_tempgrid}.
The fiducial model is repeated in each panel (dotted line) as a reference.
}
\label{fig_siggrid}
\end{figure}

We next explore the effect of changing the disk mass accretion rate and
the viscosity parameter $\alpha$. Figures \ref{fig_tempgrid} and \ref{fig_siggrid}
show the resulting disk structure
for $\Mdot=10^{-9}$, $10^{-8}$, and $10^{-7} \ \MSUNYR$, and $\alpha=0.001$,
0.01, and 0.1, for
a fixed central star with $M_*=0.5 \ \MSUN$, $R_*=2 \ \RSUN$ and $T_*=4000 \ K$.
The panels are organized in such a way that the surface density of the disk increases
towards the bottom and the right of the figure, as $\Mdot/\alpha$ increases.

Figure \ref{fig_tempgrid} shows the radial distribution of characteristic temperatures:
$T_c$, $T_0$, $T_{phot}$, and $T_{vis}$ (defined in \S \ref{sec_estruc}).
The lower the mass accretion rate,  the smaller the contribution of viscous dissipation as
a heating mechanism of the disk photosphere, which can be seen by comparing $T_{vis}$ and
$T_{phot}$.  For $\Mdot \lesssim 10^{-8} \ \MSUNYR$, the
disk photospheric temperature distribution is dominated by stellar irradiation (upper panels).

The behavior of the central temperature in the inner disk depends upon the relative
importance of viscous dissipation and irradiation
and upon the optical depth of the disk, which in turn $\propto \Sigma \propto {\Mdot}/\alpha$.
As the optical depth increases, the ratio $T_c / T_{phot}$ increases.
In all optically thick models, the central temperature shows the plateau
at $\sim 1600 K$ due to dust sublimation, as in the fiducial model (\S 2.2).
Since viscous heating at a given radius increases with
mass accretion rate, the outer boundary of the dust sublimation region moves outwards with
higher $\Mdot$.

For $\Mdot=10^{-7} \ \MSUNYR$ and $\alpha=0.001$, which corresponds to the densest disk model
shown in Figure \ref{fig_tempgrid},
the midplane temperature becomes very high, reaching  $T_c \sim 10000-30000 \ \K$ at $R
< 0.05 \AU$. The  dominant opacity source
at this high temperature range is the ionization of H,  He and metals, and the
disk probably is subject to thermal instabilities (Kawazoe \& Mineshige 1993; Bell \& Lin 1994).

In all models, irradiation heating enhanced by the flaring of the surface dominates in
the outer disk. The radius where irradiation begins to dominate 
($T_{irr} > T_{vis}$) increases
as ${\Mdot}/\alpha$ increases.
In the lowest surface density cases we have calculated,
$(\Mdot/10^{-8})/(\alpha/0.01)=0.01$ and $0.1$
(left upper corner in Figure \ref{fig_tempgrid}),
the temperature distribution flattens out at large radii.
In this  regions, the disk becomes optically thin to both
its own radiation and the stellar radiation. In this
case,
a fraction of direct and diffuse stellar radiation
is able to penetrate the disk and reach the midplane, so $\Gamma_{irr}(z=0) \ne
0$ (c.f. \S 2.2).

Figure \ref{fig_tempgrid} shows
the surface temperature inversion discussed in
\S 2.2, in many models over a wide range of radii ($T_0 > T_{phot}$ for $R \gtrsim 0.03 \ \AU$).
The upper layer temperature is approximately the
temperature of optically thin dust heated by stellar radiation only
geometrically diluted (e.g., Calvet et al. 1992; CG; DCCL) and it is independent of
$\Mdot$ and $\alpha$, being sensitive to dust properties, the effective temperature and
luminosity of
the central star, and other cooling ingredients different from dust (CO, H$_2$O, etc).

Figure \ref{fig_siggrid} shows the mass surface density of the different models,
with the fiducial $\Sigma(R)$ repeated in each panel as a reference.
>From this plot it is clear that $\Sigma$ scales as $\Mdot/\alpha$.
As mentioned above, the surface density of a steady $\alpha$-disk
is $\Sigma \sim \Mdot/\alpha T_c$, but in the outer
regions  the midplane temperature  is almost independent
of $\Mdot$ and $\alpha$, since irradiation is the dominant heating mechanism,
The disk mass, given in Table 1 for three different disk radii,
depends on the mass surface density
of the outer annuli, and
scales roughly  as $M_d \sim (R_d/100 \ \AU) (\Mdot/10^{-8} \
\MSUNYR) (0.01/\alpha) \MSUN$.

Regions in the disk are unstable to axisymmetric gravitational perturbations
if the Toomre parameter
\be
Q = {c_s \Omega \over \pi G \Sigma}
\en
(Toomre 1964) is less than unity.
It has been suggested that disks cannot maintain such regions because they will
be unstable to non-axisymmetric gravitational perturbations which effectively transfer
angular momentum, causing rapid accretion (e.g., Pringle 1981).
The most unstable regions in our models occur at large radii.
Table 1 shows the radius $R_Q$ beyond which $Q < 1$.
The most critical case is the disk model with $\Mdot=10^{-7} \ \MSUNYR$ and $\alpha=0.001$,
which is gravitationally unstable for $R > 14 \ \AU$.

\section{Comparison with Observations}

\subsection{Calculation of the Model Emission}
\label{sec_radtrans}

Given the disk structure and assuming an inclination
angle $i$  between the disk axis and  the line of sight,
we integrate the radiative transfer equation
through the disk along rays parallel to the line of sight,
in a grid of points at the plane of the sky.
To calculate the disk thermal emission
we integrate
\be
{d I_{\nu}^{therm} \over dZ} = -\kappa_\nu \rho B_\nu e^{-\tau_\nu(Z)} \ \ ,
\en
\be
{d \tau_\nu \over dZ} = -\chi_\nu \rho \ \ ,
\en
where $Z$ is the coordinate along the ray which is zero at the plane of the sky and
increases towards the observer,
$\tau_\nu(Z)$ is the monochromatic optical depth,
$B_\nu$ is the Planck function evaluated
at the local temperature, $\kappa_\nu$ is the true absorption coefficient,
and $\chi_\nu$ is the total opacity including absorption and scattering.

The contribution of the stellar radiation scattered by the disk is calculated
assuming single and isotropic scattering, so we integrate the transfer equation
\be
{d I_\nu^{scatt} \over dZ} = - \sigma_\nu W(r) B_\nu(T_*) \exp [-\tau_{\nu,rad} -\tau_\nu(Z)]\ \
,
\label{eq_intscatt}
\en
where $\sigma_\nu$ is the scattering coefficient, $W(r)$ is the geometric dilution
factor of the
stellar radiation which reaches radial distance $r$,
$B_\nu(T_*)$ is the Planck function evaluated at the stellar effective
temperature,
and $\tau_{\nu,rad}$ is the optical depth in the radial direction, between the
star
(assumed as a point-like source) and the point $r$ in the disk.
For each wavelength we calculate $\tau_{\nu,rad}(R,z)$ in cylindrical coordinates
$R,z$ once, and use this array to interpolate during the integration of equation
(\ref{eq_intscatt}).
We calculate the emergent intensity of the disk thermal radiation at  57
wavelengths between $0.55 \ \mu m$ and
$20 \ \cm$, and since  the intensity of the stellar radiation peaks around $1 \
\mu m$,
we calculate the scattered light contribution only at 13
wavelengths between $0.55 \ \mu$ and $4.65 \ \mu m$.

The intensity
emerging at each ray is crossing the
plane of the sky is
\be
I_\nu = I_\nu^{therm}+I_\nu^{scatt} + I_\nu^* e^{-\tau_\nu^*} ,
\en
where
$I_\nu^*$ is the stellar intensity, which is added if the ray intersects
the stellar disk,
and $\tau_\nu^*$ is the optical depth towards the central star produced by the disk.

Finally, we calculate the fluxes and images of
disks with arbitrary orientations and
at different wavelengths, convolving with appropriate Gaussian instrumental responses.
More details about this procedure and useful integration limits
can be found in D'Alessio (1996).
 We take the observed SED of V819 Tau which is a weak emission T Tauri Star (WTTS),
scaled to 1 $\LSUN$, as the typical central star
for calculating the system SED.

\subsection{SEDs}

\subsubsection{The Median Observed SED}
\label{sec_obs}

The principal observational constraints we use in this paper come from the SEDs of
T Tauri stars in the Taurus-Auriga molecular cloud, taken from the compilation of
KH95.
Taurus may not be the most representative region of star formation, but it is
nearby,
well-studied, and not heavily extincted, so that its stellar population is fairly
well
characterized (e.g., Hartmann et al. 1991; Gomez et al. 1993; Brice\~no et al.
1997, 1998).

Analysis of the SED of any particular source is complicated by a number of
effects.
One of the principal concerns must be the presence of binary companions,
especially
infrared bright-companions, such as in T Tau (Ghez et al. 1991), which can alter
the interpretation of the spectrum dramatically.  In some cases the near-infrared
spectra can be decomposed between the objects (e.g., Simon et al. 1992, 1995; Leinert et
al. 1993);
however, the IRAS fluxes, which are crucial to understanding the (cool) outer disk
structure, cannot similarly be separated into individual components because
of poor spatial resolution.  Beyond this, there are many other factors -
variability,
uncertainty in extinction corrections, IRAS sensitivity limits, observational
errors, effects of (generally unknown) inclination which render
the fitting of an individual SED non-unique.

For these reasons we have estimated a {\it median} TTS SED, with a measure of the
typical range of SED properties, for comparison with the disk models.
To do this we first selected K5-M2 stars from KH95 to isolate
a reasonably restricted range of stellar effective temperature
(this also happens to include a large fraction of the known T Tauri stars in
Taurus).
We eliminated the WTTS, to avoid including objects
which have little or no infrared excess emission from disks.
We also eliminated a few objects for which the data is inadequate, or
which may have infrared excesses dominated by envelope rather than
disk emission (e.g., Calvet et al. 1994).  Finally, we eliminated objects
for which the strong optical veiling makes the intrinsic system colors
uncertain, and therefore have very uncertain extinctions.
These criteria resulted in a final list of 39 objects for further study,
as listed in Table 2. As described in KH95, we adopt the luminosity estimate from the reddening-corrected
J magnitude, $L_J$, as the best estimate of the stellar luminosity (Table 2).

We then corrected the individual SEDs for extinction, using the values given in
KH95,
and normalized all the SEDs at $1.6 \mu$m.  This normalization was motivated by
the
expectation (as confirmed below) that most of the disk heating
in typical TTS is from absorbing light from the central star,
not by intrinsic accretion energy generation, in which case
the disk luminosity should scale with the stellar luminosity.
For this sample of stars, which have nearly the same effective temperature, it is
sufficient
to normalize in the H photometric band, near the peak of the stellar SED.
Adopting a near-infrared rather than optical wavelength for normalization also
helps
minimize the effects of errors in extinction corrections.

\begin{figure}
\plotone{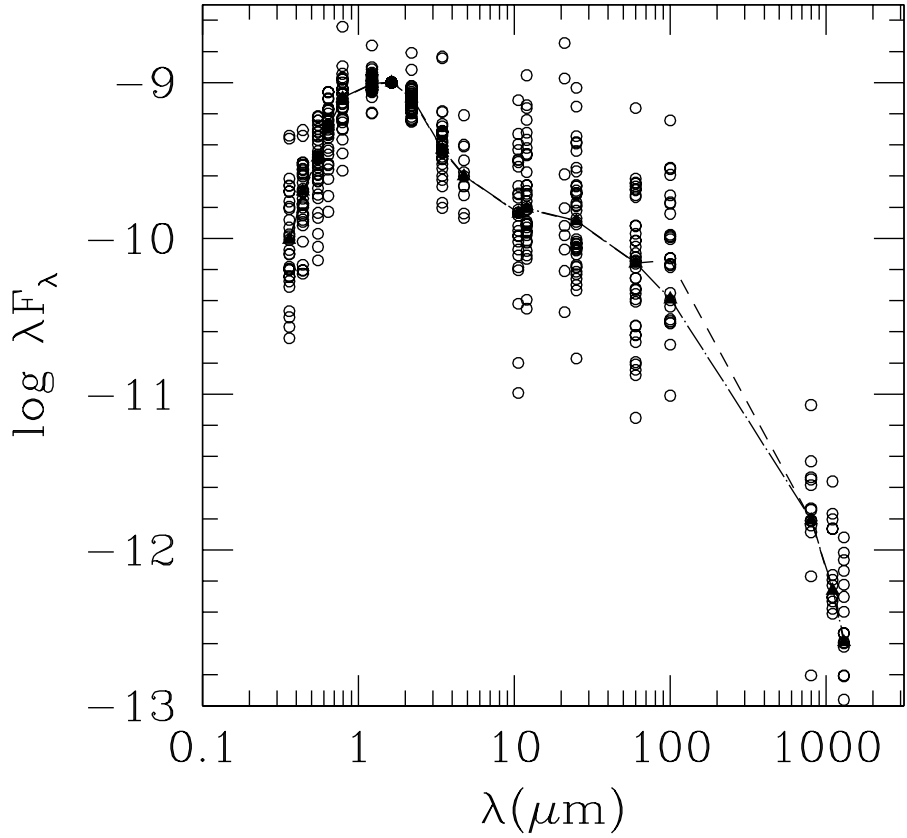}
\caption{Observed fluxes of Taurus-Auriga pre-main sequence sources
 (KH95) normalized at $\lambda=1.6 \ \mu m$ and median SED. The curves are the
median with upper limits at 100 $\mu$m (dashed line) and the median without upper limits
(dot-dashed line).}
\label{fig_median}
\end{figure}

The median fluxes and quartile limits resulting from this procedure
are given in Table 3, and the median fluxes
along with the individual observations normalized at $1.6 \mu$m
are illustrated in Figure \ref{fig_median}.  Substantial scatter
is observed at each wavelength, but there is a clearly defined trend encompassing
most objects.  The dot-dashed line is the median SED, constructed independently at each
wavelength.  This median is straightforwardly constructed except at $\lambda = 100
\mu$m,
where 13 of the 39 objects are undetected, and at mm wavelengths,
where there is very large incompleteness.  
Figure \ref{fig_median} shows 
the two estimates of the median, calculated with and without 
the upper limits at 100 $\mu$m; the difference
at this wavelength between the two estimates 
is comparable to the quartile scatter in
the median.
In the subsequent comparison between the observed median and the models
we adopt the 100 $\mu$m value which allows for incompleteness (i.e.,
calculated with upper limits).
 We emphasize that the mm-wave points are meant to be indicative
rather than definitive;
these fluxes are not used as constraints on disk models in this paper because we
do
not consider changing dust opacities (see below).

This is not the only way of developing a median SED.
We have tried normalizing at the J and K photometric bands, but the results
are essentially the same.
A crucial parameter of our models is the ratio of the disk brightness to the
stellar
brightness, and so normalization at the peak of the stellar SED
rather than at some wavelength dominated by disk emission seems most appropriate.

In principle, the median flux approach might not yield the correct SED spectral
indices as a function of wavelength if the underlying SEDs are not smooth.  In
practice,
this does not seem to be a problem.  The spectral indices derived from
our median spectrum agree well with the average spectral indices derived by KH95
by straight
averaging of spectral slopes without normalization.

It might be objected that our method of normalization (basically to the stellar
luminosity)
introduces additional error in cases where the accretion luminosity is not
negligible. In such cases, the absolute values of $L_{acc}$ and $L_*$ are important.
However, it can be seen from Table 2 that most
of the objects selected have quite similar stellar luminosities; half of the
sample
have $-0.27 \leq {\rm log} L_*/\LSUN \leq -0.03$, so that normalization introduces
relatively
small shifts in the data for many objects.  We think that the advantages of our
procedure in establishing
a well-determined, stable benchmark SED to test models against outweigh the
disadvantages
of giving individual differences reduced weight, especially because irradiation heating
probably dominates accretion energy release in most of the sample.
In any event, since we do not include continuum stars which are likely to have
high
accretion rates, neither our sample nor our method are well-suited to an
exploration of
the extreme range of properties among T Tauri disks.

\subsubsection{Fiducial model SED}
\label{sec_sedfiducial}

\begin{figure}
\plotone{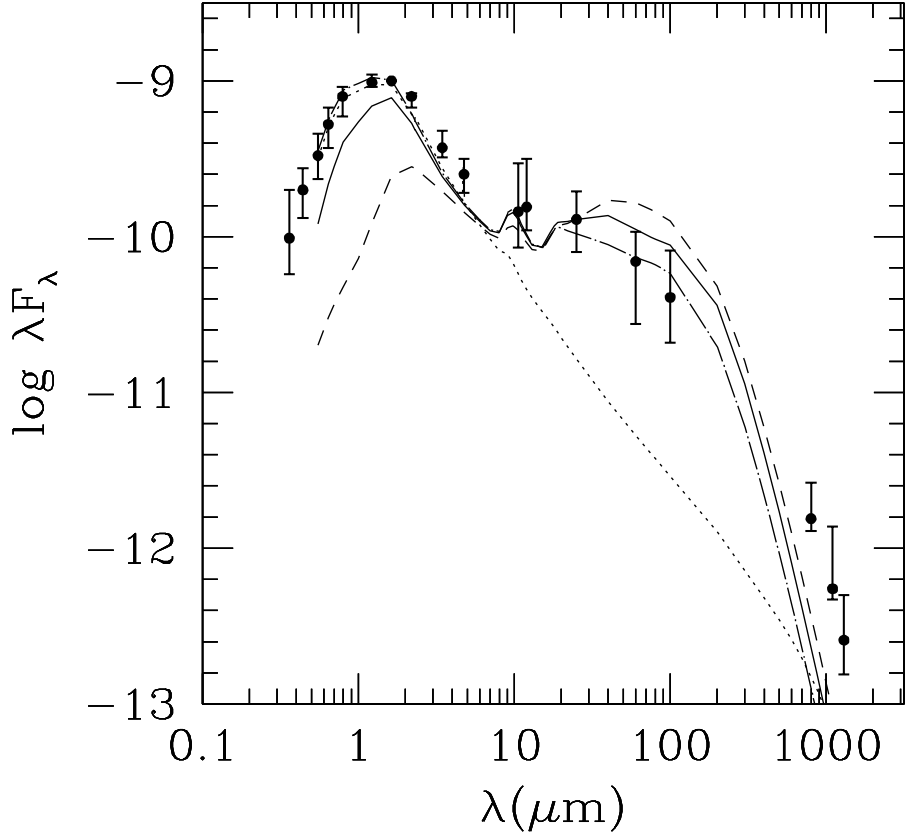}
\caption{SED of the fiducial model at an inclination angle   $i=60^\circ$ relative to the line of sight,
 for three disk radii $R_d=$ 30 AU (dot-dashed line), 100  AU (solid line),  and
300 AU (dashed line).
The median observed SED (points) and quartiles (error bars) and the disk model irradiated as a flat
disk (dotted line) are also shown.
}
\label{fig_modelmedian}
\end{figure}

Figure \ref{fig_modelmedian} shows the median observed SED
compared to the SEDs of the  fiducial model and of a flat disk with the same stellar
and disk
parameters, assuming an inclination angle of $i=60^\circ$.
Fluxes are presented for the fiducial model truncated at three different outer
radii,
30, 100, and 300 AU.

It is evident that the fiducial model, viewed at the median inclination
$i=60^\circ$ expected for a random distribution of orientations,
is roughly consistent with the SEDs of many T Tauri
stars from near-infrared wavelengths out to around 100 $\mu$m.
However, the model fluxes are slightly too low around $\lambda \sim 3 \mu$m,
and somewhat too high in the $20 - 100 \mu$m range, when compared with
the median SED, though the latter problem may be reduced if disk radii
are smaller than the typical 100 AU estimate.

The largest systematic problem of the fiducial model occurs at mm wavelengths,
where
the model fluxes are too low by a factor of 10 (for $\lambda > 1 \ mm$).
In contrast with the disk flux at shorter wavelengths, which is dominated
by optically-thick regions,
much of the disk emission in the mm range is produced
by optically thin regions, and so is strongly dependent upon the
dust opacity at long wavelengths.  The typical practice in modeling mm emission
of T Tauri disks is to adjust the dust opacity to larger values than resulting
from
the Draine and Lee (1984) calculations.  A frequently-used estimate (e.g., Beckwith \etal 1990)
would increase the 1 mm opacity by a factor of 10 over what we are using here,
and such an opacity increase would bring the model fluxes into much better
agreement
with observations.  We defer discussion of matching the long-wavelength fluxes to
the next paper in this series, where we will consider the effect of changing
dust opacities.

The comparison between the fiducial model and the flat disk model shows that
flaring
is relatively unimportant in determining the flux in the $2 < \lambda < 6 \ \mu$m
wavelength
range.  To estimate the required additional heating of the inner
region, we made the experiment of increasing the irradiation flux
by a factor of 2.5 for $R < 0.1 \ \AU$; this increases the disk photospheric
temperature by a factor of
1.2, and gives rise to  a spectrum identical to the observed median
SED in the range $1.22 < \lambda < 6 \ \mu$m.  This required large increase in
heating
suggests that reducing the dust albedo will not explain the discrepancy. Alternatively,
an increase of $\Mdot$ of a factor $ \sim 3$ would produce the requiring
heating (see \S \ref{sec_sedgrid}).

The far-infrared flux increases as the outer disk radius increases because the
flaring
results in more light from the central star to be intercepted by the disk.
The large far-IR excesses are a consequence of the substantial vertical
thicknesses
of the disk models. The corresponding result is that more lines of sight to the
central star are strongly extincted by the disk.
For the fiducial parameters, an outer disk radius of 300 AU results
in substantial extinction of the star by the disk; at $i = 60^{\circ}$, the star
would
appear to be extincted by approximately $A_V \sim 4$, as shown in Figure
\ref{fig_modelmedian}.
Although in principle this extinction would be removed by the reddening correction
used in constructing the model SEDs, this value is rather large in comparison
with the extinctions adopted for most Taurus Class II sources (Table 2).

To illustrate the disk extinction in more detail,
we calculated surfaces $z(A_V)$ of constant $A_V$ along the line of sight to the
star.  It is convenient to present these results in terms of a critical
angle $i_c$, such that $\cos i_c = \mu_c = z(A_V)/[z(A_V)^2+{R_d}^2]^{1/2}$.  Thus,
$\mu_c(A_V,{R_d})$ is the cosine of the inclination angle at which
an observer would view the central star through $A_V$ magnitudes of visual
extinction for a disk of outer radius ${R_d}$.  For inclination angles
greater than $ i_c =\cos^{-1} \mu_c$, the extinction is larger, and vice versa.

\begin{figure}
\plotone{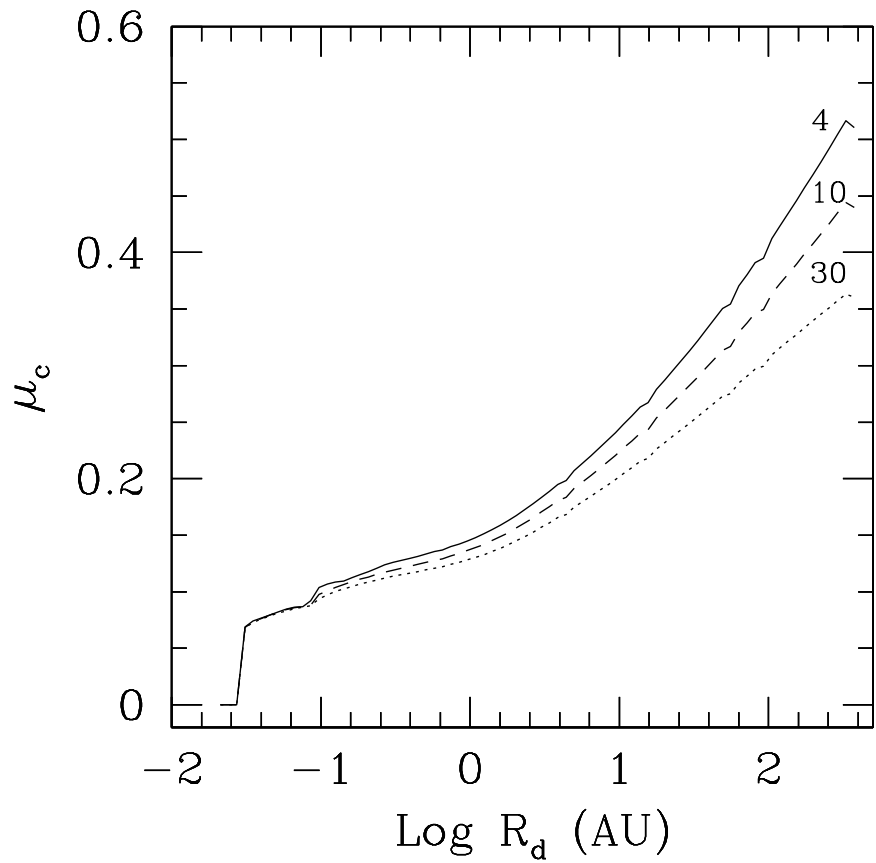}
\caption{Cosine of critical inclination angle for the fiducial model, and $A_V=4$
(solid line),
$A_V=10$ (dashed line) and $A_V=30$ (dotted line), as a function of disk radius.}
\label{fig_oculta}
\end{figure}

Figure \ref{fig_oculta} shows $\mu_c$ for $A_V=4, 10$ and $30$ for the fiducial
model.  The curves show why the models with outer radii $\leq 100$~AU exhibited
only
small extinction toward the central star at $i = 60^{\circ}$ ($\mu = 0.5$),
while the 300 AU outer radius model heavily extincted the star.
The fiducial model has a substantial vertical thickness
(cf. Figure \ref{fig_estruc}b); the ``disk photosphere''
that an observer might define at visual wavelengths would be quite
thick, with a height of nearly half the cylindrical radius at 100 AU.

\begin{figure}
\plotone{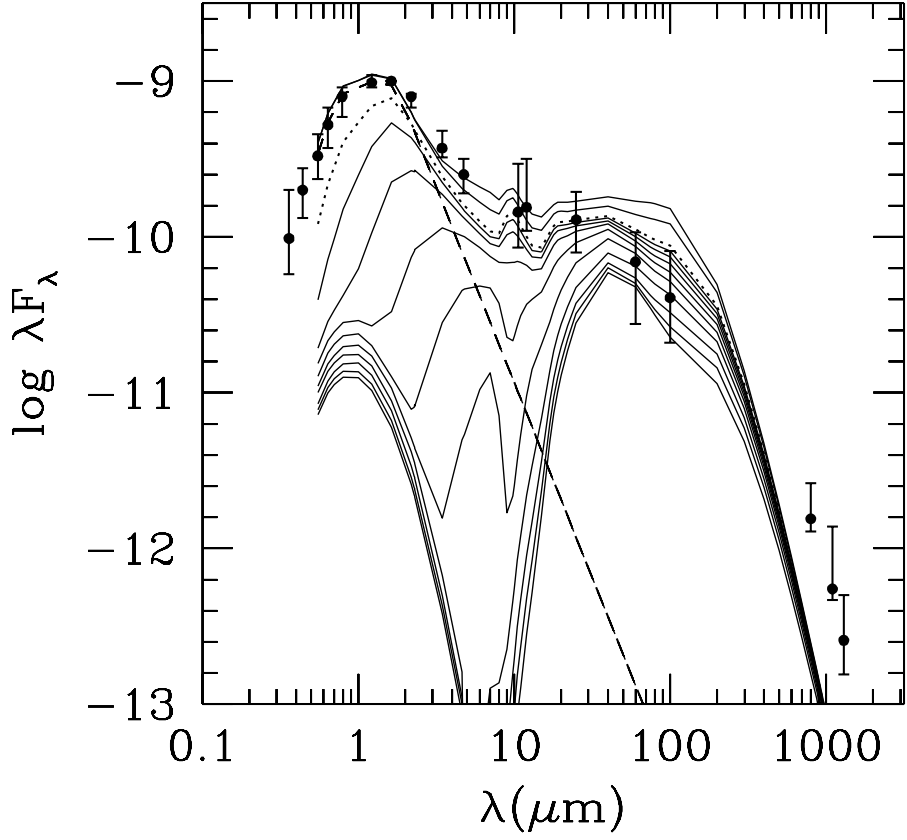}
\caption{SEDs of the fiducial disk model with $R_d=100 \ \AU$ and different
inclination angles, $\cos(i)=$ 0, 0.1, 0.15, 0.2, 0.25, 0.3, 0.35, 0.4, 0.45, 0.5,
0.75 and 1.0 (from bottom to top). As a reference, the SED corresponding to $\cos(i)=0.5^\circ$ is shown
with dotted line.
The median observed SED is also plotted (points and error bars).
}
\label{fig_angles}
\end{figure}

The large vertical thickness of the fiducial disk model
implies that the central star would be heavily extincted at a wide range of
inclination angles.  To illustrate this in detail,
Figure \ref{fig_angles} shows SEDs of our fiducial model as a function of inclination,
fixing $R_d=100 \ \AU$.  
As inclination increases,
the star first becomes heavily reddened.
At large inclination,
$\mu < 0.3$, the SED becomes separated into two distinct peaks, with the
short-wavelength
emission resulting from stellar light scattered by the disk into the line of sight.

Large vertical thickness and corresponding large extinction towards
the star for a large range of inclination angles are also found
by CG and Chiang \& Goldreich (1998), who assumed
that the dust and gas are well mixed as we do in this work. However, 
the disk thickness and the resulting infrared emission
are larger in the Chiang \& Goldreich models than in ours.
The difference arises for  two reasons. First, 
we include a finite albedo.
Second, as discussed in \S \ref{sec_estruc}, the disk atmosphere in our calculations
has a range of temperatures (similarly to a stellar 
atmosphere), and the temperature where most of
the emission of the upper atmospheric levels
is produced is lower than the single temperature
CG assume for these upper layers.
 
Conversely, the isothermal disk models of Miyake \& Nakagawa (1995) 
exhibited thinner disk structure than found here. 
However, even with this property, Miyake \& Nakagawa argued that the SEDs of a few T Tauri
stars could be explained only if their disks were geometrically flat, which could
be attributed to dust settling toward the disk midplane. 

One other property to note is the presence of emission in
the 10 $\mu m$ silicate band in the low inclinations disk
models; this band is produced by the temperature inversion in
the disk atmosphere (Calvet et al. 1991, 1992), which is more conspicuous in
the SED of flared disk models than in the flat disk.
Both, the flat and the flared models have a temperature inversion in the vertical
direction
(Calvet et al. 1991, 1992), but the smaller the angle between
the incidence direction and the disk normal, the larger the  {\it vertical }
optical depth of the atmospheric layer where the largest fraction of stellar
energy is
deposited 
and the more important its contribution to the disk SED. 
As inclination increases,  a very strong silicate absorption
feature develops (as also is shown by 
Chiang \& Goldreich 1998, hereafter CG2).

\subsubsection{SED dependence on $\Mdot$ and $\alpha$}
\label{sec_sedgrid}

The models presented in \S \ref{sec_magrid} cover a range of $\Mdot$ compatible with
observations of veiling in CTTS (Valenti, Basri \& Johns 1993;
Hartigan, Edwards \& Ghandour 1995; Gullbring \etal 1998).
In this section we show how the SEDs depend on $\Mdot$ and $\alpha$ to see if
different combinations of these parameters give a better fit to the median SED than our fiducial model.

\begin{figure}
\plotone{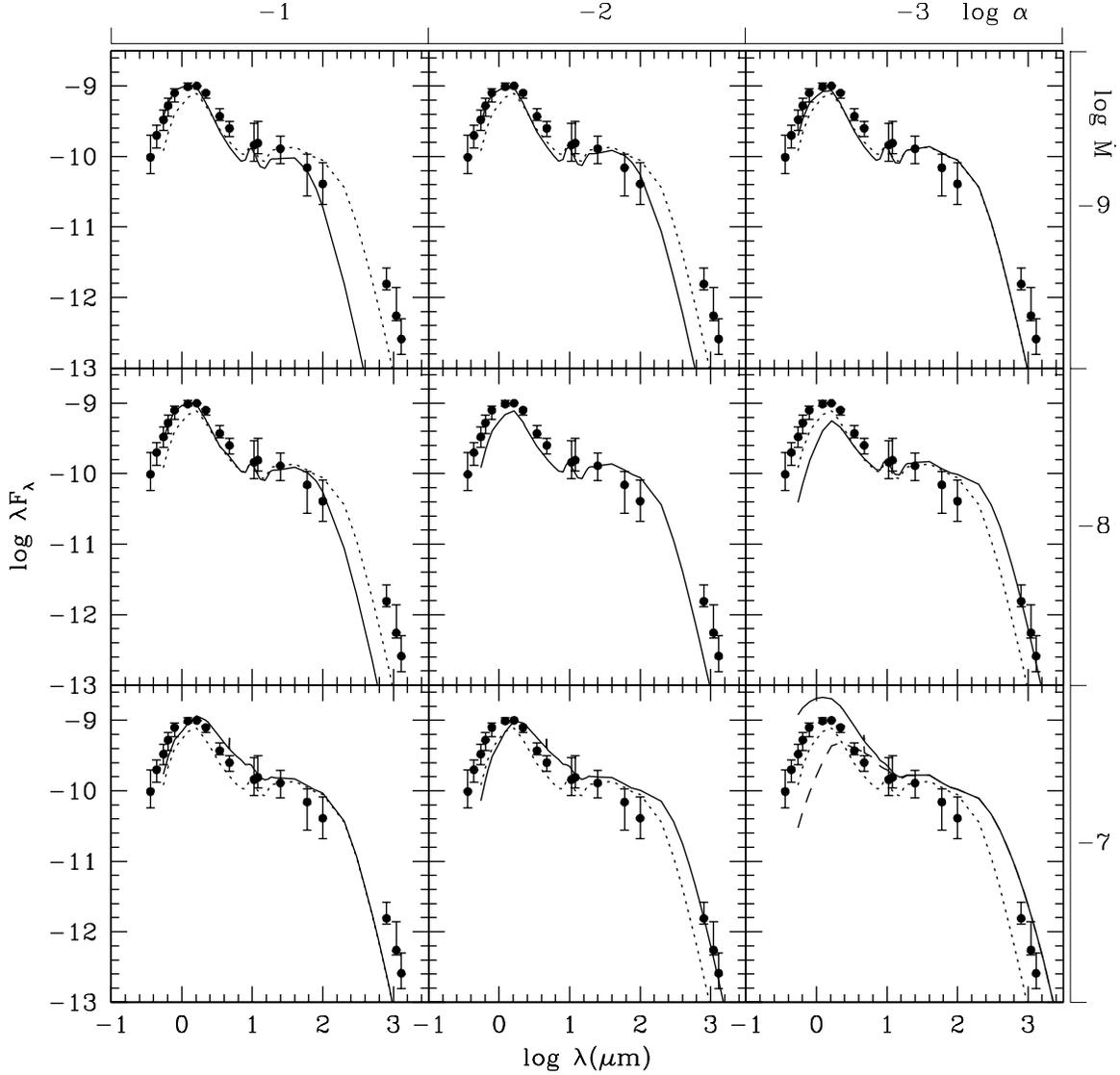}
\caption{SEDs of the same models shown in Figure \ref{fig_tempgrid}, for
$i=60^\circ$ and $R_d=100 \ \AU$,
compared with the median of the observations and quartiles (points and error bars,
respectively).
 The SED of the fiducial model is repeated in each panel (dotted line) as a
reference. In the case of $\Mdot=10^{-7} \ \MSUNYR$ and $\alpha=0.001$, the high flux
at short-wavelengths is due to the emission of the inner wall at the hole radius $R_{hole}=3 R_*$.
This inner region has a very high temperature as can be seen in Figure \ref{fig_tempgrid}, which
probably corresponds to a thermally unstable solution. The SED of a disk model with the same
parameters but a $R_{hole}=4 R_*$ does not show the high optical-near IR emission.
}
\label{fig_sedgrid}
\end{figure}

Figure \ref{fig_sedgrid} shows SEDs of irradiated accretion disks with
the same parameters as in \S \ref{sec_magrid}, namely,
$\Mdot=10^{-9}$, $10^{-8}$, and $10^{-7} \ \MSUNYR$, and $\alpha=0.001$,
0.01, and 0.1, at an inclination $i=60^\circ$.
The near-IR emission ($\sim 2 - 5 \mu$m) is dominated by optically
thick regions at $R \lesssim 1 \ \AU$ (see \S \ref{sec_magrid}).
The difference between the models with $\Mdot=10^{-9}$ and $10^{-8} \ \MSUNYR$
(upper panels in Figure \ref{fig_sedgrid}) is small because irradiation
is the dominant heating source at low accretion rates.
For these models the near IR flux (at $i = 60^{\circ}$) is smaller than the median
(approximately at the level of the lower quartiles).
On the other hand, accretion energy is important in enhancing the
optically thick emission for the $\Mdot=10^{-7} \ \MSUNYR$, and so these models
exhibit near IR fluxes larger than the median and the upper quartiles.

We find that an accretion rate of $\Mdot=3 \times 10^{-8} \ \MSUNYR$
fit the median observed SED in the 2-8 $\lambda$ spectral range reasonably well.
This is slightly larger than the median accretion
rate of $\Mdot \sim 1 \times 10^{-8} \ \MSUNYR$
found by Gullbring et al. (1998).
Note that the actual median stellar luminosity is $L_* \sim 10^{-0.13} \ \LSUN$, so
the required mass accretion rate would be $\Mdot \sim 10^{0.34} \times 10^{-8} \sim 2 \times 10^{-8} \ \MSUNYR$.
 The agreement is reasonably satisfactory
considering that there may be systematic errors in accretion rates
at the factor of two level. Moreover, the
spread between the quartile fluxes is consistent with a random
distribution of inclinations and a quartile spread of order
$\sim \pm 3$ in accretion rates consistent with observations (Hartmann et al. 1998).

The mid- to far-infrared fluxes ($10 \mu m \le \lambda \le 100 \ \mu m$)
are much less sensitive to the accretion rate because of the importance
of irradiation in outer disk regions.
In general, the models in the grid also show the same excess in this
wavelength region relative to the median SED that the fiducial model
exhibits.  The exception is the lowest accretion rate, largest-$\alpha$
model; in this case the $10-100 \mu$m flux is reduced because the very low
mass disk becomes optically thin.  However, these model parameters do not seem to
represent a good solution to the excess mid- to far-IR fluxes, because
the mm-wave fluxes become extremely small, and it would require an enormous
increase in the long-wavelength opacity to reconcile the model with observations.

The flux in the mm range tends to scale as the disk mass
(e.g., Beckwith \etal 1990), which in our grid of fixed radius models scales as $\Sigma \propto \Mdot/\alpha$.
Figure \ref{fig_sedgrid} shows that models with an $\Mdot/\alpha$ larger by a factor of 10
with respect to the fiducial model (i.e., $\Mdot=10^{-8} \ \MSUNYR$ and $\alpha=0.001$
or $\Mdot=10^{-7} \ \MSUNYR$ and $\alpha=0.01$) can account for the mm emission
without changing the dust opacity (cf. discussion in previous subsection).
However, the masses of these disks are $M_d \sim 0.13-0.14 \ \MSUN$ (Table 1),
not far from the limit for gravitational stability (Table 1; \S 3.2.2).
In addition, the high $\Mdot$ of these disks
results in near infrared fluxes much higher
than the median SED.

At short wavelengths, the extinction of the stellar radiation
produced by the disk is shown by the difference
between the model fluxes and the median.  As $\Mdot/\alpha$ increases,
so does the disk density and thus the $A_V$ towards the central star.
To illustrate how the extinction depends on the disk parameters,
Table 4 lists the cosine of the critical inclination angle, $\mu_c$,
for  $A_V=4$ and $R_d=100 \ \AU$.
The excess of extinction of the central star radiation (cf. Table 4) and
the excess of flux in the range  $25 - 100 \ \mu m$ are a consequence
of the large vertical thickness of the disk models.

\subsubsection{Images}

\begin{figure}
\plotone{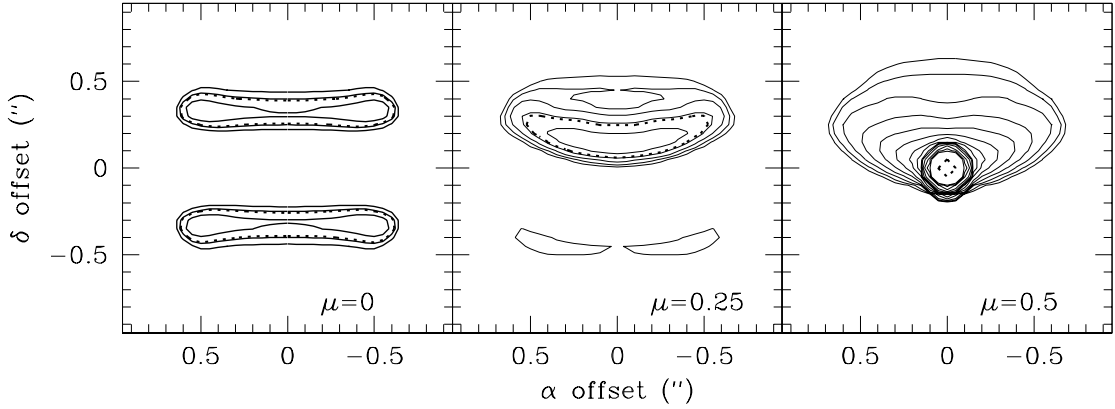}
\caption{Images of the fiducial disk model for different inclination
angles $i=90^\circ$ ($\mu=0$), $i=75.5^\circ$ ($\mu=0.25$) and $i=60^\circ$ ($\mu=0.5$).
The maximum contour level is $I_{max} = 11.56 \ mJY/beam$, the minimum is
$I_{min}=0.03 \ mJY/beam$, and each contour level is a factor 1.58 larger
than the previous one (this corresponds to variations of 0.5 mag). The dotted contour
 corresponds to the half maximum brightness of each image. The images are calculated assuming a
distance of $d=140$ pc to the Taurus molecular cloud.}
\label{fig_images}
\end{figure}

Figure \ref{fig_images} shows images of
the fiducial disk model for
inclination angles such that $cos i = \mu=0, \ 0.15, \ 0.3 $ and $0.4$, and a disk radius
$R_d=100 \ \AU$.
The images  are calculated at
$\lambda=1 \ \mu$m, convolving with a Gaussian point-spread function
with a full width half maximum of 0.1 arcseconds (i.e., 14  AU at a
distance $d=140$ pc). The contour levels are separated by a factor 1.58, which is
approximately 0.5 magnitudes.
For $\mu < 0.3$ the flux from the star is obscured by the intervening
disk material, and the scattered light intensity relative to the peak
intensity is high. For $\mu > 0.3$ the stellar radiation dominates the image,
and the light scattered by the disk is not detectable by contrast.

The image of the edge-on disk ($\mu=0$) shows two elongated
reflection nebulosities, almost parallel, separated by a dark lane.
The apparent thickness of the disk, measured as the distance along the
polar axis between
the center of the elongated nebulae below and above the disk plane,
is $\Delta z_{app} = 0.64" = 89 \AU$, and the aspect ratio of the
image, as defined by thickness divided by diameter, is $\sim 0.5$.  It is similar in shape to, but larger in aspect
ratio than, the disk model
shown by Whitney \& Hartmann (1992) (cf. their Figure 6), which
has an aspect ratio of $\sim 0.3$.

There are images of two known edge-on disks in Taurus with which
we can compare the model predictions. The image of HH30 (Burrows \etal 1996)
shows a clearly flared geometry.
This geometry is not so apparent in the theoretical image because
the extinction due to the dust in the outer disk is large,
even at large heights, resulting in a very thick dark lane which hides the
curvature of the irradiation surface.
The HH30 disk can be detected out to a radius $R_d=250 \ \AU$, which is larger than
our fiducial model; however, increasing the radius of the disk model does not change the shape
of the image, which still is not as flared as observed
(as shown by Whitney \& Hartmann 1992, cf. their Figure 7).

Burrows \etal (1996) found that the (isothermal) disk model which
produces the best fit of the observed HH30 images has a scale height
of 15.5 AU at $R=100 \ \AU$ (similar values are found by Wood \etal 1998).
We compare this height to the theoretical local scale height
(defined in \S \ref{sec_method})
evaluated at the upper atmospheric layers, since
these are the layers where most of the scattering is taking place.
The scale height determined from fitting the scattered
light is factor of 2 smaller than the
gas scale height of the upper atmosphere of our fiducial model, calculated using the temperature  $T_0$.
This suggests that  the region of the atmosphere where most of the scattering is happening
is colder than the upper atmospheric layers of the fiducial model,
supporting the idea of a less  flared irradiation surface.

HK Tau/c (Stapelfeldt \etal 1998; Koresko 1998) has a shape closer to the
image of the edge-on model, i.e., two parallel reflection nebulae
separated by a dark lane. The apparent thickness of this disk
is $\Delta z_{app} \sim 29 \pm 3$ AU (Koresko 1998) and the thickness of the
dark lane is 0.13 times the apparent width of the nebulae (Stapelfeldt \etal 1998).
In the case of the edge-on model, the relative thickness of the dark lane is $\sim
3$ times the observed one.
Again, we conclude that the disk model is too thick to explain
the observations.
The disk model that Stapelfeldt \etal (1998)  proposed as the best fit for HK Tau/c
has a scale height of 3.8 AU at $R=$ 50 AU.
The scale height of the disk model
evaluated at $T_{irr}$ is a factor of 2 higher,
and it is even larger at $T_0$,
suggesting again that the disk has to be thinner and colder than the model.

In summary, the thickness of the predicted dark lane in our edge-on fiducial model
is larger than what is observed in HK Tau/c, and the overall disk structure
is inconsistent with the observations of HH30, which also probably require
a thinner absorption layer, suggesting that our fiducial model is
a factor of $\sim 2$ too thick in the vertical direction.

\section{Discussion}

\subsection{Edge-on Disks and Class I sources}
\label{sec_edge}

If T Tauri disks have substantial geometrical thicknesses, then
a sizeable number of objects should be viewed edge-on, through the disk.
Many heavily-extincted young stellar objects could be
edge-on disk systems, detected only as infrared sources.
These considerations may suggest that some of the
Class I infrared sources, usually interpreted as protostars surrounded by infalling envelopes
(e.g., Adams \etal 1987; Butner et al. 1991;  Kenyon, Calvet \& Hartmann 1993) 
could be instead simply T Tauri stars
viewed through their disks (CG2).
For example,
CG2 presented a SED calculation for a disk system viewed at an intermediate
angle which agrees reasonably well with that of an individual Class I source in Taurus.

\begin{figure}
\plotone{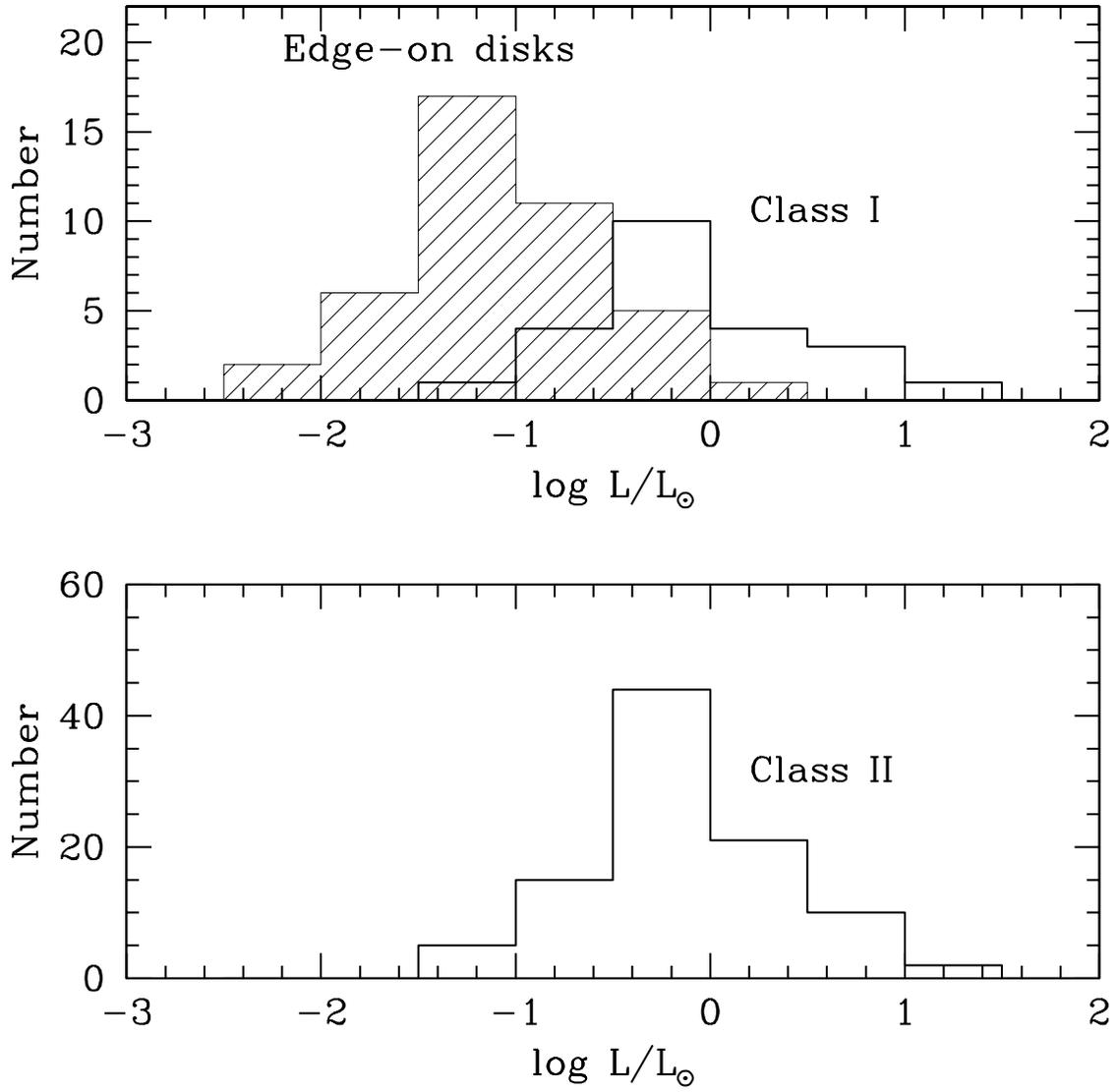}
\caption{Luminosity distributions of Class I and II Sources in Taurus (KH95), and edge-on disk models
(see text)}
\label{fig_lum}
\end{figure}

As in the case of T Tauri stars, in considering Class I sources
it is helpful to analyze the properties of an
entire population of objects rather than simply
fitting an individual source.  One clear prediction is that
edge-on systems should appear less luminous than they really are,
since the disk emission is reduced at large inclinations
and the central star becomes heavily extincted (Figure \ref{fig_angles}).
If Class I objects were simply Class II (face-on disk) systems viewed edge-on,
then the Class I sources should appear to have systematically
lower luminosities than Class II objects.
Yet the luminosity distributions of Class I and Class II sources in Taurus
(taken from KH95) are nearly identical (Figure \ref{fig_lum}).
To make a quantitative test, we have constructed
a predicted luminosity distribution for edge-on disk systems
as follows.
For a given stellar luminosity, $L_*$, we calculate a frequency distribution
of observed luminosities for the star+disk system assuming
a random distribution of inclinations.
We scaled the fluxes of the fiducial
model to $L_*$ to obtain the luminosity of the
star+disk system at a given inclination.
We restrict the range of
inclinations to $0 \le \mu \le 0.3$, which corresponds to
SEDs with $\lambda F_{\lambda}$ increasing with $\lambda$, 
as expected for Class I objects.
Finally, we convolved this frequency distribution
with the observed distribution of stellar luminosities
of Class II sources in Taurus.
The luminosity distribution resulting from this
procedure, shown in Figure \ref{fig_lum},
indicates that Class I sources are brighter than 
expected from edge-on disks,
suggesting that only a modest fraction of these sources are likely to be edge-on disk systems.
The luminosity distribution resulting from this
procedure, shown in Figure \ref{fig_lum},
differs substantially from that of the currently recognized Class I sources in Taurus,
suggesting that only a modest fraction of these sources are likely to be edge-on disk systems.

\begin{figure}
\plotone{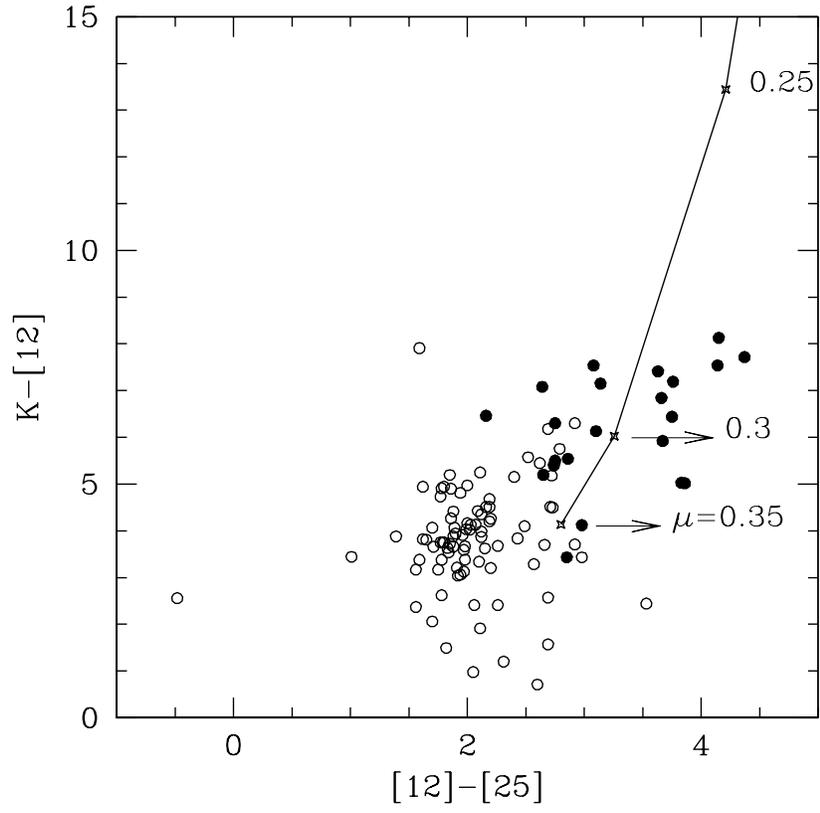}
\caption{Color-color diagram for K-[12] and [12]-[25]. Observed colors are indicated
by open circles (Class II Sources) and filled circles (Class I Sources), and are taken
from KH95. The solid line represents the colors of the fiducial model for inclination angles
$i > 69.5 ^\circ$ ($\mu < 0.35$).}
\label{fig_color}
\end{figure}

Another problem with interpreting Class I sources as edge-on disks
is that the observed SEDs do not agree with the models over a
significant range of inclinations.
In particular, our models tend to predict extremely large silicate
absorption, which is not observed in general (Kenyon, Calvet \& Hartmann 1993).
This can be illustrated by the color-color diagram  $K-[12]$  vs. $[12]-[25]$,  shown in Figure
\ref{fig_color}. The circles represent the observed colors of Class II (open circles)
and Class I (dark circles) sources in Taurus (KH95), the line corresponds
to the colors of the fiducial model for $\mu < 0.35$.
The predicted color $K-[12]$, which quantifies the slope of the SED between the
near and mid IR, increases with the inclination angle and for $ \mu \lesssim 0.3$
becomes larger than the maximum observed $K-[12]$ for a Class I source.
Only disks within a narrow range of inclination angles  would have the IR colors
and bolometric luminosities consistent with observed Class I sources.
Because the probability of observing
an object in a given range of inclination is proportional to the range in $\mu$,
the model predicts a much larger number of sources with
$K-[12]$ and $[12]-[25]$ colors much redder than observed.

In summary,
while our models suggest, in agreement with CG2, that some Class I sources
may be edge-on disk systems,
most Class I sources are unlikely to be edge-on disk systems because
they are not systematically underluminous in comparison with
Class II (face-on disk) systems, and because their infrared colors and SEDs
are inconsistent with disk model predictions for all but a narrow range of inclinations.

\subsection{``Missing'' sources}

The model calculations shown in Figure \ref{fig_color} indicate that approximately
30\% of all T Tauri disk systems should have colors and infrared SEDs which
are not characteristic of known objects.  In Taurus, with a total
Class II T Tauri population of $\sim 100$ objects (KH95), this would correspond to about
$\sim 40$ systems with peculiar SEDs.  While this appears to be strongly inconsistent
with observations, one must also consider selection effects.  In particular,
since edge-on systems are fainter, one must consider whether surveys have
systematically missed these sources.

The first question we address is whether edge-on disk systems
would be detectable as IRAS sources.  As above, we scaled the edge-on disk models
to the stellar luminosity distribution of Class II sources, but consider
the flux distribution at 60 $\mu$m instead of the luminosity distribution.
We find that approximately 80 \% of the edge-on disks should be detectable,
using a flux limit of 0.5 Jy at 60 microns as adopted in the
survey of faint Taurus sources by Kenyon et al. (1994b).
Scaling from the known Class II sources in Taurus, this would suggest that
there should be approximately 34 edge-on sources in the region which
should be detected from IRAS surveys.

\begin{figure}
\plotone{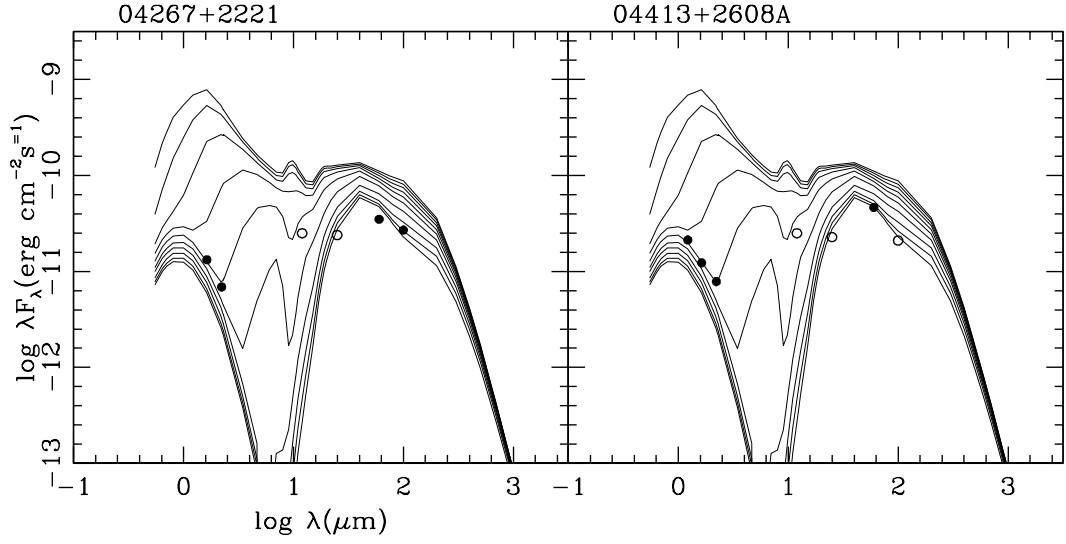}
\caption{Examples of candidates for edge-on disks from the sample of Kenyon
et al. (1994b). SEDs of the fiducial disk model for different inclination angles
are shown for reference. open circles correspond to upper limits.}
\label{fig_missing}
\end{figure}

To test this hypothesis, we use the results of
Kenyon et al. (1994b), who searched the IRAS Serendipitous Survey
Catalog for previously unidentified point sources with
fluxes exceeding 0.5 Jy at 60 microns, adding some other
candidates with fluxes above 0.3 Jy at 25 or 60 microns
provided they were detected in at least two bands from
the SSC and the PSC2 catalogs.
We selected objects from Kenyon et al. (1994b) which
are not galaxies as found by near-IR imaging and which
also were not classified as Class I or II sources.
There are 19 such sources, out of which we estimated
8 candidates by comparison with theoretical SEDs at high inclinations.
Figure \ref{fig_missing}  shows two of
these candidates
for edge-on systems: I04267+2221 and I04413+2608A.
However, this estimate is likely to
be a generous upper limit to the number of possible edge-on sources, since
there is a lack of multiwavelength data for most of these objects.
Some of these sources could still be galaxies (9) as mentioned by Kenyon et al. (1994b).
In addition, not all objects with near-IR and far-IR peaks
are necessarily edge-on disks.  It is known that many young binary
systems have greatly differing extinctions to the individual stars,
and this multicomponent SED could mimic that of an edge-on disk.
But even counting all these candidates and including the
$\sim$ 10 Class I objects that overlap the predicted
distribution of edge-on systems
(Figure \ref{fig_lum}), we end up with 19
candidates, a factor of $\sim$ 2 lower than expected.

In summary, our disk models with
well-mixed gas and dust  appear to predict a larger number of
low-luminosity, but detectable, highly reddened objects than
consistent with current surveys of Taurus.

\section{Conclusions}

We have presented models of irradiated accretion disks with well-mixed
ISM dust and compared them with observations representative
of the entire population of Classical T Tauri stars in Taurus.  We find
that our models can explain the near-infrared fluxes of the median
SED of T Tauri stars with a disk accretion
rate consistent with the mean value estimated by
Gullbring \etal (1997). However, our
models appear to be too geometrically thick at large radii, as suggested
by: (a) larger far-infrared disk emission than the typical
SEDs of T Tauri stars; (b) scattered light images with wider dark
dust lanes than observed in two objects in Taurus; and (c) model predictions
of large numbers of T Tauri stars hidden by their edge-on disks which
contradict current surveys.

Solar nebula theory predicts that dust settling and coagulation could occur on
timescales that are comparable to or shorter than T Tauri lifetimes,
(e.g., Hayashi, Nakazawa \& Nakagawa 1985; Weidenschilling \& Cuzzi 1993; Weidenschilling 1997).
Because dust dominates the disk opacity, settling can strongly reduce the
geometrical thickness of the disk, and may be required to explain the observations
of T Tauri stars (Miyake \& Nakagawa 1995).
We will consider this possibility in a subsequent paper.

This work was supported in part by the Visiting Scientist Program of the Smithsonian
Institution, by NASA grant NAG5-4282, by Academia Mexicana de Ciencia and
Fundaci\'on M\'exico-EEUU para la Ciencia, by DGAPA/UNAM and CONACyT, M\'exico.

\newpage

\begin{center}
\begin{tabular}[h]{ l c c c c r}
\hline
\multicolumn{6}{c}{\bf TABLE 1} \\
\multicolumn{6}{c}{\bf MODEL DISK MASSES AND TOOMRE RADIUS} \\
\hline
\hline
$log \Mdot$\tablenotemark{a} & $log \alpha$ & $M_d(R_d=30 \ \AU)$\tablenotemark{b} & $M_d(R_d=100 \ \AU)$ & $M_d(R_d=300 \ \AU)$ & $R_Q$\tablenotemark{c}  \\
\hline
-9      &       -1      &$4.6 \times 10^{-5}$ &$1.0 \times 10^{-4}$ &$2.2 \times 10^{-4}$       &$>340$ \\
-9      &       -2      &$4.6 \times 10^{-4}$ &$1.5 \times 10^{-3}$ &$3.2 \times 10^{-3}$  &$>340$ \\
-9      &       -3      &$4.4 \times 10^{-3}$ &$1.4 \times 10^{-2}$ &$4.0 \times 10^{-2}$ &$>340$ \\
-8      &       -1      &$4.6 \times 10^{-4}$ &$1.5 \times 10^{-3}$ &$3.2 \times 10^{-3}$ &$>340$ \\
-8      &       -2      &$4.3 \times10^{-3}$  &$1.4 \times 10^{-2}$ &$4.1 \times 10^{-2}$ &$>340$  \\
-8      &       -3      &$4.0 \times 10^{-2}$ &$1.3 \times 10^{-2}$ &$4.0 \times 10^{-1}$ &$132$ \\
-7      &       -1      &$4.1 \times 10^{-3}$ &$1.4 \times 10^{-2}$ &$4.0 \times 10^{-2}$ &$>340$ \\
-7      &       -2      &$3.7 \times 10^{-2}$ &$1.3 \times 10^{-1}$ &$4.0 \times 10^{-1}$ &$173$ \\
-7      &       -3      &$2.8 \times 10^{-1}$ &$1.2 \times 10^{0}$  &$3.9 \times 10^{0}$ &$14$ \\
\hline
\tablenotetext{a}{\scriptsize {Mass accretion rate in $\MSUNYR$}}
\tablenotetext{b}{\scriptsize {Masses in $\Mdot$}}
\tablenotetext{c}{\scriptsize {Radius in AU}}
\end{tabular}
\end{center}

\newpage

\begin{center}
\begin{tabular}[h]{ l r r }
\hline
\multicolumn{3}{c}{\bf TABLE 2} \\
\multicolumn{3}{c}{\bf STARS USED FOR MEDIAN SED} \\
\hline
Object & $A_V$ & log $L_J/\LSUN$ \\
\hline
HO Tau       &1.15   & -0.72 \\
HN Tau       &0.53   & -0.66 \\
DM Tau       &0.00   & -0.60 \\
GO Tau       &1.21   & -0.55 \\
FM Tau       &0.71   & -0.49 \\
FS Tau       &1.89   & -0.49 \\
DP Tau       &1.49   & -0.39 \\
IP Tau       &0.25   & -0.37 \\
CY Tau       &0.09   & -0.33 \\
V710 Tau     &0.90   & -0.27 \\
DI Tau       &0.78   & -0.21 \\
DS Tau       &0.31   & -0.19 \\
IQ Tau       &1.27   & -0.19 \\
CoKu Tau/3   &3.35   & -0.18 \\
DH Tau       &1.27   & -0.17 \\
HK Tau       &2.39   & -0.17 \\
DD Tau       &0.78   & -0.14 \\
DQ Tau       &0.99   & -0.14 \\
AA Tau       &0.50   & -0.13 \\
JH 112       &3.32   & -0.13 \\
LkCa 15      &0.65   & -0.13 \\
V819 Tau     &1.40   & -0.09 \\
GM Aur       &0.15   & -0.08 \\
GI Tau       &0.90   & -0.07 \\
CI Tau       &1.83   & -0.06 \\
FY Tau       &3.56   & -0.05 \\
DN Tau       &0.50   & -0.04 \\
V955 Tau     &2.76   & -0.03 \\
BP Tau       &0.50   & -0.02 \\
GK Tau       &0.90   &  0.07 \\
DO Tau       &2.70   &  0.08 \\
Haro 6-37    &2.17   &  0.11 \\
LkHa 332/G1  &3.25   &  0.15 \\
GG Tau       &0.78   &  0.18 \\
DF Tau       &0.22   &  0.20 \\
UZ Tau E     &1.52   &  0.20 \\
UY Aur       &1.40   &  0.30 \\
FV Tau       &4.84   &  0.34 \\
V807 Tau     &0.00   &  0.61 \\
\hline
\end{tabular}
\end{center}

\newpage

\begin{center}
\begin{tabular}[h]{ l r r r}
\hline
\multicolumn{4}{c}{\bf TABLE 3} \\
\multicolumn{4}{c}{\bf MEDIAN SED AND QUARTILES} \\
\hline
Wavelength ($\mu$m) & log $\lambda F_{\lambda}$ (med) & upper & lower \\
\hline
\hline
    0.36  &-10.01   &-10.24  &-9.70        \\
    0.44  & -9.70   &-9.88   &-9.56        \\
    0.55  & -9.48   &-9.63   &-9.34        \\
    0.64  & -9.28   &-9.43   &-9.17       \\
    0.79  & -9.10   &-9.23   &-9.04       \\
     1.22  & -9.01   &-9.04   &-8.96        \\
     1.63  & -9.00   &-9.00   &-9.00       \\
     2.19  & -9.10   &-9.17   &-9.08        \\
     3.45  & -9.43   &-9.49   &-9.32        \\
     4.75  & -9.60   &-9.72   &-9.50        \\
     10.60  & -9.84  &-10.07   &-9.53        \\
     12.00  & -9.81   &-9.96  & -9.50       \\
     25.00  & -9.89   &-10.10  &-9.71        \\
     60.00 & -10.16   &-10.56  &-9.97        \\
    100.00 & -10.43   &-10.73  &-10.13       \\
     800.00  & -11.81  &-11.89  &-11.58       \\
     1100.00  & -12.26  & -12.33 & -11.86      \\
     1300.00  & -12.59  & -12.81 & -12.30     \\
\hline
\end{tabular}
\end{center}
Note. - The $100 \mu$m entry is the median flux for the
sample including upper limits; the lower quartile flux 
cannot be directly determined because of the large number
of non-detections, so the quartile spread is assumed
to be the same as when considering only detections
(see text).

\newpage
\begin{center}
\begin{tabular}[h]{ l c c r}
\hline
\multicolumn{4}{c}{\bf TABLE 4} \\
\multicolumn{4}{c}{\bf MODEL DISK CRITICAL ANGLE  } \\
\hline
\hline
$log \Mdot$ & $log \alpha$ & $\mu_c(A_V=4)$ &$ i_c$\\
\hline
-9      &       -1      &0.25 &75.5\\
-9      &       -2      &0.34 &70.1\\
-9      &       -3      &0.41 &65.8\\
-8      &       -1      &0.34 &70.1\\
-8      &       -2      &0.41 &65.8\\	
-8      &       -3      &0.46 &62.6\\
-7      &       -1      &0.41 &65.8\\
-7      &       -2      &0.46 &62.6\\
-7      &       -3      &0.50 &60.0\\
\hline
\end{tabular}
\end{center}

\newpage

\end{document}